\documentclass[
aip,
nofootinbib]{revtex4-1}
\usepackage{graphicx}
\usepackage{pstricks-add}
\usepackage{amsmath}
\usepackage{amssymb}
\usepackage{braket}
\usepackage{mathrsfs}
\usepackage{hyperref}
\usepackage{natbib} 
\usepackage{listings}
\usepackage{tikz}
\usepackage{simpler-wick}
\newif\ifstartedinmathmode
\newcommand\encircled[1]{%
  \relax\ifmmode\startedinmathmodetrue\else\startedinmathmodefalse\fi%
  \tikz[baseline,anchor=base]{%
  \node[draw,circle,outer sep=0pt,inner sep=.2ex]
    {\ifstartedinmathmode$#1$\else#1\fi};}%
}
\begin{document}
\title{A Thermofield-based Multilayer Multiconfigurational Time-Dependent Hartree Approach to Non-Adiabatic Quantum Dynamics at Finite Temperature}

\author{Eric W. Fischer}
\email{ericwfischer@posteo.de}
\affiliation{Theoretische Chemie, Institut f\"ur Chemie, Universit\"at Potsdam,
Karl-Liebknecht-Stra\ss{}e 24-25, D-14476 Potsdam-Golm, Germany}

\author{Peter Saalfrank}
\email{peter.saalfrank@uni-potsdam.de}
\affiliation{Theoretische Chemie, Institut f\"ur Chemie, Universit\"at Potsdam,
Karl-Liebknecht-Stra\ss{}e 24-25, D-14476 Potsdam-Golm, Germany}

\let\newpage\relax

\begin{abstract}
We introduce a thermofield-based formulation of the multilayer multiconfigurational time-dependent Hartree (ML-MCTDH) method to study finite temperature effects on non-adiabatic quantum dynamics from a non-stochastic, wave-function perspective. Our approach is based on the formal equivalence of bosonic many-body theory at zero temperature with doubled number of degrees of freedom and the thermal quasi-particle representation of bosonic thermofield dynamics (TFD). This equivalence allows for a transfer of bosonic many-body MCTDH as introduced by Wang and Thoss to the finite temperature framework of thermal quasi-particle TFD.
As an application, we study temperature effects on the ultrafast internal conversion dynamics in pyrazine. We show, that finite temperature effects can be efficiently accounted for in the construction of multilayer expansions of thermofield states in the framework presented herein. Further, we find our results to agree well with existing studies on the pyrazine model based on the $\rho$MCTDH method.
\end{abstract}

\let\newpage\relax
\maketitle


\newpage

\section{Introduction}
The study of quantum systems at finite temperature is a central topic in non-equilibrium quantum statistical mechanics\cite{stefanucci2013,breuer2007} and chemical physics of condensed phases\cite{nitzan2014}. The description of a quantum system at finite temperature extends wave function theory to the concept of density operators\cite{blum2012} and, therefore, naturally enhances the strong exponential scaling of the underlying Hilbert space. This ''curse of dimensionality'' is traditionally studied from the reduced perspective of (open system) density matrix theory\cite{breuer2007,blum2012} by treating only a subsystem of interest explicitly, which then interacts implicitly with the remaining degrees of freedom. An alternative approach, which explicitly includes contributions beyond the subsystem, is based on the concept of purification\cite{zwolak2004,verstraete2004,feiguin2005}. The purification ansatz maps a density operator to a wave function equivalent representation and therefore allows to benefit from numerically powerful methods as the density matrix renormalization group (DMRG)\cite{schollwoeck2011}, and more generally matrix product states (MPS)\cite{paeckel2019}, to tackle dynamics in the finite temperature regime. Closely related to the concept of purification is the theory of thermofield dynamics (TFD)\cite{takahashi1975,semenoff1983,takahashi1996}, which provides a many-body equivalent formulation of quantum statistical mechanics. In TFD, the density operator is linked to pure thermofield states evolving in time according to a Schr\"odinger type equation of motion, which replaces the Liouville-von Neumann equation on an artificially extended Hilbert space. Although TFD plays an important role in theoretical physics\cite{umezawa1982,umezawa1993,khanna2009,blasone2011}, it has only recently entered the field of chemical physics\cite{ritschel2015,reddy2015,borrelli2016,borrelli2017,gelin2017,wang2017,borrelli2018,
harsha2019a,harsha2019b,shushkov2019,borrelli2019,begusic2020,begusic2021,borrelli2021,gelin2021}, where it has been combined with tensor trains/matrix product states (TT/MPS)\cite{borrelli2016,borrelli2019,borrelli2021,gelin2021} and the multi-Davydov D2 ansatz\cite{wang2017} to tackle the ``curse of dimensionality'' issue.
\\
A well established and powerful approach in molecular quantum dynamics is the multiconfigurational time-dependent Hartree (MCTDH) method\cite{meyer1990,manthe1992,beck2000,meyer2009,meyer2012} and its multilayer extension (ML-MCTDH)\cite{wang2003,manthe2008,vendrell2011,wang2015}. The MCTDH ansatz had been initially formulated for molecular vibrational degrees of freedom but has later been successfully generalized to the treatment of fermions and bosons\cite{zanghellini2003,kato2004,nest2005,alon2008,
wang2009,cao2013,kroenke2013,manthe2017,weike2020}. {Moreover, the finite temperature regime has been accessed by directly propagating the density operator in terms of the $\rho$MCTDH approach\cite{raab1999,raab2000a,raab2000b,meyer2003,picconi2019}, which was, however, restricted to a small set of system DoFs. Additionally, wave function based stochastic sampling approaches to the evaluation of thermal ensemble averages\cite{matzkies1999,manthe2001,nest2007,lueder2010,lorenz2014} have been considered, which potentially suffer from a large number of necessary realizations to properly sample the initial thermal state rendering them costly for large systems.}
\\
In this work, we augment the MCTDH approaches to the finite temperature regime by formulating the ML-MCTDH method in the theoretical framework of thermofield dynamics, which allows us to access quantum systems at finite temperature from a {\em non-stochastic} perspective in a numerically established framework of 
{\em wave function} (rather than density matrix) theory. We show how the existing MCTDH ansatz can be transferred to TFD, by combining the second quantization representation (SQR)\cite{wang2009} of MCTDH and the thermal quasi-particle (TQP) representation of TFD\cite{takahashi1996}. The advantage of our approach is twofold: First, the numerical power of the ML-MCTDH ansatz, especially for high-dimensional problems\cite{thoss2006,wang2006,wang2007,craig2007,wang2013,wang2018}, can be employed to mitigate the scaling problem at finite temperatures. This is in contrast to $\rho$MCTDH, which has not been formulated in the multilayer framework so far. Second, the presented approach is directly applicable via the efficient (ML)-MCTDH implementation in the \textit{Heidelberg MCTDH package}\cite{heidelbergmctdh}. To test our approach, we apply it to the study of finite temperature effects in non-adiabatic quantum dynamics. In particular, we consider the ultrafast internal conversion dynamics in pyrazine at finite temperature, which has already been studied by means of the density operator formulation of MCTDH in Refs.\cite{raab1999,raab2000a}, and present results for a corresponding 24-mode system-bath model, respectively.
\vspace{0.2cm}
\\
The paper is organized as follows. In section \ref{sec.nonadiabat_thermofield}, we briefly discuss issues  of non-adiabatic quantum dynamics at finite temperature, which motivates the introduction of thermofield dynamics (TFD) and its thermal quasi-particle (TQP) representation. In section \ref{sec.mctdh_tqp}, we build on the TQP concept to formulate the multiconfigurational time-dependent Hartree ansatz in the framework of TQP and discuss its properties. In section \ref{sec.model_observables}, we introduce the thermal quasi-particle representation of the vibronic coupling model for ultrafast internal conversion in pyrazine at finite temperature in the TFD framework. In section \ref{sec.results}, we first discuss the computational performance of our ansatz, followed by an examination of temperature effects on internal conversion and linear absorption spectra for linear and bilinear vibronic coupling Hamiltonians of pyrazine and the impact of a bilinearly coupled harmonic bath on the linear model. Finally, section \ref{sec.conclusion} concludes our work.

\section{The Thermofield Approach to Quantum Statistical Dynamics}
\label{sec.nonadiabat_thermofield}
\subsection{Non-Adiabatic Molecular Quantum Dynamics at Finite Temperature}
\label{subsec.nonadiabat_statmech}
We start by recapitulating the basic theory of non-adiabatic, molecular quantum dynamics at finite temperature from a statistical quantum dynamics perspective. The time evolution of a quantum system at finite temperature is completely described by the density operator $\hat{\rho}(t)$, which obeys the Liouville-von Neumann (LvN) equation\cite{breuer2007,nitzan2014}
\begin{equation}
\dfrac{\partial}{\partial t}\,
\hat{\rho}(t)
=
-\dfrac{\text{i}}{\hbar}
\left[
\hat{H},
\hat{\rho}(t)
\right],
\hspace{0.75cm}
\hat{\rho}(t_0)
=
\hat{\rho}_0
\label{eq.liouville_von_neumann}
\end{equation}
with initial state $\hat{\rho}_0$. In a non-adiabatic context, the molecular dynamics governed by Eq.\eqref{eq.liouville_von_neumann} is generated by a vibronic Hamiltonian\cite{raab2000a} 
\begin{equation}
\hat{H}
=
\sum^{N_e}_{i=1}
\hat{H}_i
\ket{S_i}
\bra{S_i}
+
\sum^{N_e}_{i\neq j}
\hat{V}_{ij}
\ket{S_i}
\bra{S_j},
\label{eq.vibronic_hamiltonian}
\end{equation}
here given in the basis of $N_e$ diabatic electronic states $\{\ket{S_i}\}$. The first term contains the molecular vibrational Hamiltonian, $\hat{H}_i=E_i+\hat{H}_0+\hat{V}_{c,i}$, of the $i^{\text{th}}$ electronic state with electronic energy $E_i$. In the vibronic coupling model, 
 $\hat{H}_0=\sum_k \hbar \omega_k 
\left(
\hat{a}^\dagger_k
\hat{a}_k
+
\frac{1}{2}
\right)$ is the uncoupled, vibrational-mode Hamiltonian with normal-mode frequencies $\omega_k$ and normal-mode bosonic creation and annihilation operators $\hat{a}^\dagger_k$ and $\hat{a}_k$ for the $k^\mathrm{th}$ mode, respectively. The same $\hat{H}_0$ is assumed for every electronic state. Further, $\hat{V}_{c,i}$ are intra-state, inter-mode vibronic coupling operators which couple different modes $k$ and $k'$, also expressible by bosonic creation and annihilation operators (see below). The second term in the vibronic Hamiltonian Eq.\eqref{eq.vibronic_hamiltonian} contains the inter-state coupling elements $\hat{V}_{ij}$, which in general also depend on the vibrational mode operators $\{\hat{a}^\dagger_k,\hat{a}_k\}$.

The non-adiabatic quantum dynamics at moderate temperature, \textit{i.e.}, only vibrational modes are excited thermally, is obtained by solving the LvN equation \eqref{eq.liouville_von_neumann} for 
\begin{equation}
\hat{\rho}(t)
=
\sum^{N_e}_{i,j=1}
\hat{\rho}^{ij}_v(t)
\ket{S_i}
\bra{S_j}
\label{eq.vibronic_density_operator}
\end{equation}
with vibrational density operator $\hat{\rho}^{ij}_v(t)$ and  electronic states $|S_i\rangle$. In what follows, we will consider an uncorrelated initial state, $\hat{\rho}(t_0)=\hat{\rho}^0_\beta\,\ket{S_0}\bra{S_0}$, with 
\begin{equation}
\hat{\rho}^0_\beta
=
\dfrac{e^{-\beta\hat{H}_0}}{Z^0_\beta},
\hspace{0.5cm}
Z^0_\beta
=
\sum^M_{k=1}\,e^{-\beta\varepsilon_k}
\end{equation}
calculated from the normal-mode ground-state Hamiltonian $\hat{H}_0$ and corresponding eigenvalues $\{\varepsilon_k\}$. (In the pyrazine model used below, there are no intermode coupling terms for the ground state, {\em i.e.}, $\hat{V}_{c,0}=0$. An extension to vibrationally coupled initial states is straightforward). $M$ is the number of vibrational states, $\beta=(k_B T)^{-1}$, and $Z^0_\beta$ is the normal-mode partition function. A subsequent external perturbation,  here assumed to be sudden and complete, excites the system and initiates the non-adiabatic dynamics. The time-evolved density operator $\hat{\rho}(t)$ gives observables as 
\begin{equation}
\braket{\hat{O}}_\beta(t)
=
\mathrm{tr}
\left\{
\hat{\rho}(t)\,
\hat{O}
\right\},
\label{eq.thermal_average}
\end{equation}
where the trace runs over both the electronic and the vibrational degrees of freedom. 

A general issue in quantum dynamics is the unfavorable exponential scaling of the underlying Hilbert space, which is particularly severe in a density matrix/finite temperature context. Assuming a ground-state normal-mode basis expansion with $M$ vibrational states in every electronic state $i$, the vibrational density matrix $\hat{\rho}^{ij}_v(t)$ is 
\begin{equation}
\hat{\rho}^{ij}_v(t)
=
\sum^M_{\underline{n},\underline{n}^\prime}
\rho^{ij}_{\underline{n}\,\underline{n}^\prime}
\ket{\underline{n}}
\bra{\underline{n}^\prime},
\label{eq.density_operator_basis}
\end{equation}
with multi-index $\underline{n}=(n_1,\dots,n_f)$ for $f$ vibrational modes. The vibronic density operator Eq.\eqref{eq.vibronic_density_operator} in combination with Eq.\eqref{eq.density_operator_basis} suffers from exponential scaling with exponent $\sim N^2_e\,M^{2f}$, in contrast to wave function theory, which scales exponentially "only" with $\sim N_e\,M^f$. 

In this work, we approach the scaling issue of finite temperature non-adiabatic molecular quantum dynamics by formulating the multilayer multiconfigurational time-dependent Hartree (ML-MCTDH) method\cite{wang2003,manthe2008,vendrell2011} in the framework of thermofield dynamics (TFD) motivated by Ref.\cite{borrelli2016}. TFD provides a wave function-equivalent formulation of quantum statistical mechanics, which renders the numerically powerful ML-MCTDH ansatz accessible. It avoids direct propagation of density matrices, and also the propagation of $K$ wave functions as in stochastic wavepacket methods (where $K$ can be large).

\subsection{Basics of Thermofield Dynamics}
\label{subsec.basics_tfd}
We introduce the basic concepts of thermofield dynamics (TFD) from the perspective of a vibrational density operator, $\hat{\rho}_v(t)$. Our discussion is first restricted to dynamics on the electronic ground state, and generalized to a non-adiabatic framework afterwards.

In the symmetric formulation of TFD\cite{umezawa1982,takahashi1996}, the (vibrational) density operator $\hat{\rho}_v(t)$ reads 
\begin{equation}
\hat{\rho}_v(t)
=
\mathrm{tr}_{\tilde{\mathcal{H}}}
\left\{
\ket{\psi_\beta(t)}
\bra{\psi_\beta(t)}
\right\},
\label{eq.thermal_state_definition}
\end{equation}
with normalized, time-dependent (vibrational) thermofield state $\ket{\psi_\beta(t)}$, \textit{i.e.}, $\braket{\psi_\beta(t)\vert\psi_\beta(t)}=1$, defined on the thermal Fock space $\mathcal{H}_\beta=\mathcal{H}\otimes\tilde{\mathcal{H}}$. Here, $\mathcal{H}$ is the Fock space of the physical vibrational system and $\tilde{\mathcal{H}}$ is an exact copy of $\mathcal{H}$ denoted as auxiliary Fock space. In TFD, the auxiliary modes provide an artificial thermal bath, which allows to reconstruct the density operator from the thermofield states by performing the trace $\mathrm{tr}_{\tilde{\mathcal{H}}}\left\{\dots\right\}$ in Eq.\eqref{eq.thermal_state_definition} only with respect to the auxiliary vibrational subspace $\tilde{\mathcal{H}}$.\cite{takahashi1996}

The unitary time-evolution of the thermofield state $\ket{\psi_\beta(t)}$ is determined by the thermofield time-dependent Schr\"odinger equation (TF-TDSE)\cite{takahashi1996,khanna2009}
\begin{equation}
\dfrac{\partial}{\partial t}\,
\ket{\psi_\beta(t)}
=
-\dfrac{\text{i}}{\hbar}\,
\left(
\hat{H}
-
\tilde{H}
\right)
\ket{\psi_\beta(t)},
\hspace{0.75cm}
\ket{\psi_\beta(t_0)}
=
\ket{\psi_\beta}
\label{eq.thermofield_lvn}
\end{equation}
with hermitian thermofield Hamiltonian $\hat{H}-\tilde{H}=\bar{H}$ and initial state $\ket{\psi_\beta}$. The auxiliary Hamiltonian, $\tilde{H}$, is an exact copy of $\hat{H}$ and acts exclusively on the auxiliary subspace $\tilde{\mathcal{H}}$. The thermofield Hamiltonian $\bar{H}$ resembles the non-interacting physical and auxiliary vibrational modes. In symmetric TFD, thermal ensemble averages are calculated for observables $\hat{O}$ as 
\begin{equation}
\braket{\hat{O}}_\beta(t)
=
\braket{\psi_\beta(t)
\vert
\hat{O}
\vert
\psi_\beta(t)},
\label{eq.thermofield_thermal_average}
\end{equation}
which are formally equivalent to thermal expectation values as given in Eq.\eqref{eq.thermal_average}.\cite{takahashi1996} 
\\
As initial vibrational thermofield state, we consider a thermal vacuum state, \textit{i.e.}, $\ket{\psi_\beta}=\ket{\underline{0}_\beta}$, which takes for normal modes in symmetric TFD the form\cite{takahashi1996}
\begin{align}
\ket{\underline{0}_\beta}
&=
\dfrac{e^{-\beta\hat{H}_0/2}}{\sqrt{Z^0_\beta}}
\sum_{\underline{n}}
\underbrace{\prod^{f}_{k=1}
\dfrac{\left(
\hat{a}^\dagger_k
\right)^{n_k}}{\sqrt{n_k!}}
\dfrac{\left(
\tilde{a}^\dagger_k
\right)^{n_k}}{\sqrt{n_k!}}
\ket{\underline{0},\tilde{\underline{0}}}}_{=\ket{\underline{n},\tilde{\underline{n}}}},
\label{eq.thermal_vacuum_system}
\end{align}
where both physical and auxiliary modes share the same summation index $\underline{n}$ and the tilde sign multi-index $\underline{\tilde{n}}=(\tilde{n}_1,\dots,\tilde{n}_f)$ is only present to distinguish both types of DoF. Further, $\ket{\underline{0},\tilde{\underline{0}}}$ is a direct product state, and $\hat{H}_0$ and $Z^0_\beta$ are normal-mode Hamiltonian and partition function as introduced above. The thermal equilibrium state $\hat{\rho}^0_\beta$ is recovered from $\ket{\underline{0}_\beta}$ via the definition in Eq.\eqref{eq.thermal_state_definition}, \textit{i.e.}, $\hat{\rho}^0_\beta=\mathrm{tr}_{\mathcal{\tilde{H}}}\{\ket{\underline{0}_\beta}\bra{\underline{0}_\beta}\}$. Finally, in Eq.\eqref{eq.thermal_vacuum_system}, we introduced the auxiliary normal-mode creation operator, $\tilde{a}^\dagger_k$, which, together with an annihilation operator, $\tilde{a}_k$, generates auxiliary number states $\ket{\tilde{n}_k}$ in analogy to the physical operators $\hat{a}^\dagger_k,\hat{a}_k$ and physical number states $\ket{n_k}$ (see Appendix A for details).

A problematic aspect of $\ket{\underline{0}_\beta}$ as initial state is related to its highly non-trivial character due to the presence of strongly correlated physical and auxiliary vibrational basis states. In order to circumvent this difficulty, we introduce the thermal quasi-particle (TQP) representation of TFD\cite{takahashi1996}, which provides an advantageous representation of the thermal vacuum state $\ket{\underline{0}_\beta}$, that turns out to be particularly suited for the formulation of the ML-MCTDH ansatz in the TFD framework.

\subsection{The Thermal Quasi-Particle Representation of TFD}
\label{subsec.tqp_tfd}
From the perspective of this work, a particular useful representation of $\ket{\underline{0}_\beta}$ identifies the latter as vacuum state for a set of new normal-mode operators $\{\hat{b}_k,\tilde{b}_k\}$, acting as
\begin{equation}
\hat{b}_k\ket{0^{(k)}_\beta}
=
\tilde{b}_k\ket{0^{(k)}_\beta}
=
0  \quad  .
\label{eq.thermal_vacuum_tqp}
\end{equation}
The thermal quasi-particle (TQP) operators $\hat{b}_k$ and $\tilde{b}_k$ characterize $\ket{0^{(k)}_\beta}$ as a two-mode thermal vacuum state such that the highly entangled nature of $\ket{0^{(k)}_\beta}$ is only implicitly present. {In particular in the TQP representation, the multi-mode thermal vacuum state $\ket{\underline{0}_\beta}$ factorizes into a simple Hartree product, 
\begin{equation}
\ket{\underline{0}_\beta}
=
\ket{0^{(1)}_\beta}
\ket{0^{(2)}_\beta}
\dots
\ket{0^{(f)}_\beta} \quad ,
\label{eq.hartree_product_thermal_vacuum}
\end{equation}
of $f$ two-mode thermal vacuum states, respectively, which is by definition uncorrelated. The normal-mode thermal quasi-particle operators, $\hat{b}_k$ and $\tilde{b}_k$, are defined via the thermal Bogoliubov transformation (TBT)\cite{takahashi1996}, which provides a unitary rotation on the thermal Fock space that mixes physical and auxiliary operators (see Appendix A for details) at fixed inverse temperature $\beta$. We note, that the simplicity of the multi-mode thermal vacuum state in the TQP representation (\textit{cf.} Eq.\eqref{eq.hartree_product_thermal_vacuum}) comes at the cost of interactions between physical and auxiliary DoFs in the respective TQP thermofield Hamiltonian $\bar{H}_\beta$, which have to be accounted for in the time-evolution of the TQP thermofield states.}

\subsection{Non-Adiabatic Quantum Dynamics in Thermal Quasi-Particle TFD}
\label{eq.nonadiabat_tqp_tfd}
We now reformulate the non-adiabatic problem presented in Sec.(\ref{subsec.nonadiabat_statmech}) in the framework of thermal quasi-particle TFD, where we follow arguments by Borrelli and Gelin\cite{borrelli2016}. The vibronic thermofield Hamiltonian takes the form
\begin{equation}
\bar{H}
=
\sum^{N_e}_{i=1}
\underbrace{\left(
E_i
+
\hat{H}_i
-
\tilde{H}_0
\right)}_{=\bar{H}_i}
\ket{S_i}
\bra{S_i}
+
\sum^{N_e}_{i\neq j}
\hat{V}_{ij}
\ket{S_i}
\bra{S_j} \quad ,
\label{eq.tfd_vibronic_hamiltonian}
\end{equation}
where only the uncoupled normal-modes are treated from a thermofield perspective via the auxiliary Hamiltonian $\tilde{H}_0$, leaving other terms of the vibronic coupling Hamiltonian unaffected. As shown in Ref.\cite{borrelli2016}, Eq.\eqref{eq.tfd_vibronic_hamiltonian} can be justified due to the different energy scales of electronic and vibrational excitations. The TQP representation, $\bar{H}_\beta$, $\bar{H}^i_\beta$ and $\hat{V}^{ij}_\beta$, of operators $\bar{H}$, $\bar{H}_i$ and $\hat{V}_{ij}$ is obtained via the inverse TBT (\textit{cf.} Appendix A, Eqs.\eqref{eq.inverse_tbt_phys} and \eqref{eq.inverse_tbt_aux}). In Sec.(\ref{sec.model_observables}), we provide an explicit example of a TQP thermofield Hamiltonian for the vibronic coupling mode Hamiltonian of pyrazine.

Further, we introduce the vibronic thermofield state, $\ket{\Psi_\beta(t)}$, given in the basis of diabatic electronic states as 
\begin{equation}
\ket{\Psi_\beta(t)}
=
\sum^{N_e}_{i=1}
\ket{\psi^i_\beta(t)}
\ket{S_i},
\label{eq.vibronic_thermofield_state}
\end{equation}
which relates with Eq.\eqref{eq.thermal_state_definition} to the vibronic density operator, \textit{i.e.}, $\hat{\rho}(t)=\mathrm{tr}_{\tilde{\mathcal{H}}}\{\ket{\Psi_\beta(t)}\bra{\Psi_\beta(t)}\}$. In the TQP representation, the thermofield-TDSE \eqref{eq.thermofield_lvn} subsequently takes the form
\begin{equation}
\dfrac{\partial}{\partial t}
\ket{\Psi_\beta(t)}
=
-
\dfrac{\text{i}}{\hbar}
\bar{H}_\beta
\ket{\Psi_\beta(t)},
\hspace{0.75cm}
\ket{\Psi_\beta(t_0)}
=
\ket{S_i}
\ket{\underline{0}_\beta},
\end{equation}
and can be solved by expanding $\ket{\Psi_\beta(t)}$ as
\begin{equation}
\ket{\Psi_\beta(t)}
=
\sum^{N_e}_{i=1}
\underbrace{\left(
\sum^M_{\underline{n},\underline{m}}
C^i_{\underline{n}\,\underline{m}}(\beta)
\ket{\underline{n}_\beta,\tilde{\underline{m}}_\beta}
\right)}_{=\ket{\psi^i_\beta(t)}}
\ket{S_i} \quad ,
\label{eq.thermofield_state_basis_expansion}
\end{equation}
with TQP number states $\ket{\underline{n}_\beta,\tilde{\underline{m}}_\beta}$ (\textit{cf.} Appendix A), analogously to the standard approach in wave function theory. This expansion scales as $\sim N_e\,M^{2f}$, \textit{i.e.}, only linear in the number of electronic states $N_e$ opposed to the vibronic density operator\cite{borrelli2016}, but still strongly exponentially in the vibrational DoFs. In order to mitigate the ``curse of dimensionality'' problem, we introduce in the following the multilayer multiconfigurational time-dependent Hartree (ML-MCTDH) ansatz for thermofield states $\ket{\Psi_\beta(t)}$ formulated in the TQP representation of the thermal Fock space. 

\section{A Multiconfigurational Time-Dependent Hartree Approach to Thermal Quasi-Particle TFD}
\label{sec.mctdh_tqp}
The MCTDH approach for TQP thermofield states $\ket{\Psi_\beta(t)}$ is based on the formal analogy between the thermal quasi-particle representation of TFD at fixed $\beta$ and a $2f$-dimensional bosonic many-body problem at zero temperature. This formal equivalence allows for a transfer of the bosonic many-body formulation of MCTDH, \textit{i.e.}, MCTDH-SQR as introduced by Wang and Thoss\cite{wang2009}, to the framework of bosonic thermal quasi-particle TFD. 

The multiconfigurational time-dependent Hartree expansion of a general vibronic thermofield state $\ket{\Psi_\beta(t)}$ is given by
\small{
\begin{equation}
\ket{\Psi_\beta(t)}
=
\sum^{N_e}_{k=1}
\sum^{n_1,\dots,n_f}_{j_1,\dots,j_f}
\sum^{m_1,\dots,m_f}_{i_1,\dots,i_f}
A^{(1)}_{j_1,\dots,j_f,i_1,\dots,i_f,k}(t,\beta)
\left(
\prod^f_{\kappa,\tau=1}
\ket{\varphi^{(1,\kappa)}_{j_\kappa}(t,\beta)}\,
\ket{\tilde{\varphi}^{(1,\tau)}_{i_\tau}(t,\beta)}
\right)
\ket{S_k},
\label{eq.mctdh_tqp_expansion}
\end{equation}}
\normalsize
with time- and temperature-dependent tensorial coefficients $A^{(1)}_{j_1,\dots,j_f,i_1,\dots,i_f}(t,\beta)$ , orthonormal thermal single particle functions (tSPFs) $\ket{\varphi^{(1,\kappa)}_{j_\kappa}(t,\beta)},\ket{\tilde{\varphi}^{(1,\tau)}_{i_\tau}(t,\beta)}$ and electronic states $\ket{S_k}$, respectively. The equations-of-motion (EoM) for coefficients and tSPFs are obtained by employing the Dirac-Frenkel time-dependent variational principle at fixed $\beta$ as
\begin{equation}
\braket{
\delta\Psi_\beta(t)
\vert
\text{i}\hbar
\dfrac{\partial}{\partial t}
-
\bar{H}_\beta
\vert
\Psi_\beta(t)}
=
0,\,
\hspace{0.5cm}
\beta
=
\mathrm{const.},
\label{eq.tqp_dirac_frenkel}
\end{equation}
in the TQP representation of $\mathcal{H}_\beta$, which leads to a set of coupled, nonlinear differential equations identical to the standard MCTDH approach\cite{manthe1992,beck2000,meyer2012}. In order to make the thermal quasi-particle TFD formulation explicit, we consider the notion of thermal SPFs and abbreviate our approach as MCTDH-TQP. The expansion of $\ket{\Psi_\beta(t)}$ in Eq.\eqref{eq.mctdh_tqp_expansion} follows a formal decomposition of the thermal Fock space $\mathcal{H}_\beta(2f)$ as a tensor product of $2f$ thermal single-mode subspaces given by
\begin{equation}
\mathcal{H}_\beta
=
\underbrace{
\mathcal{H}^{(1)}(1)
\otimes
\dots
\otimes
\mathcal{H}^{(f)}(1)
}_{=\mathcal{H}}
\\
\otimes
\underbrace{
\tilde{\mathcal{H}}^{(1)}(1)
\otimes
\dots
\otimes
\tilde{\mathcal{H}}^{(f)}(1)
}_{=\tilde{\mathcal{H}}},
\label{eq.thermfock_single_modes}
\end{equation} 
where we follow the notation of Wang and Thoss\cite{wang2009}. The tSPFs $\ket{\varphi^{(1,\kappa)}_{j_\kappa}(t,\beta)}$ and $\ket{\tilde{\varphi}^{(1,\tau)}_{i_\tau}(t,\beta)}$ can be subsequently expanded in a primitive basis of single-mode TQP number states $\{\ket{n^{(l_\kappa)}_\beta}\}$ and $\{\ket{\tilde{m}^{(l_\tau)}_\beta}\}$ (\textit{cf.} Appendix A, Eq.\eqref{eq.thermal_tqp_numberstates}), spanning the respective subspaces and share for, $\kappa=\tau$, a common two-mode thermal vacuum state $\ket{0^{(\kappa)}_\beta}$. Further, the decomposition in Eq.\eqref{eq.thermfock_single_modes} is not unique and a straightforward generalization to multi-mode subspaces\cite{wang2009} as 
\begin{equation}
\mathcal{H}_\beta
=
\underbrace{
\mathcal{H}^{(1)}(d_1)
\otimes
\dots
\otimes
\mathcal{H}^{(D)}(d_{D})
}_{=\mathcal{H}}
\otimes
\underbrace{
\tilde{\mathcal{H}}^{(1)}(d^\prime_1)
\otimes
\dots
\otimes
\tilde{\mathcal{H}}^{(D^\prime)}(d^\prime_{D^\prime})
}_{=\mathcal{\tilde{H}}}
\label{eq.thermfock_combined_modes}
\end{equation} 
with, $\sum^D_\kappa d_\kappa=\sum^{D^\prime}_\tau d^\prime_\tau=f$, leads to the concept of combined modes\cite{meyer2012}. In particular, mixed multi-mode subspaces combining physical and auxiliary DoFs are possible and turn out to be advantageous for the study of high-temperature regimes as discussed below. 
\\
Finally, instead of truncating the expansion of a multi-mode tSPF via a primitive basis expansion on the corresponding, potentially mixed, multi-mode subspace, one might expand it in a new basis of time-dependent tSPFs. This approach adds additional layers to the MCTDH expansion and results in the numerically powerful multilayer formulation of the MCTDH method (ML-MCTDH)\cite{wang2003,manthe2008,vendrell2011,wang2015}, which is particularly well suited for high-dimensional problems and therefore beneficial for the strong exponential scaling of the thermofield approach in its ML-MCTDH-TQP formulation.

\section{Model Hamiltonian and Observables}
\label{sec.model_observables}
We apply the MCTDH-TQP method to finite temperature effects on ultrafast internal conversion dynamics in a well-studied pyrazine model\cite{worth1998,raabworth1999}, which was previously treated within the density matrix formulation of MCTDH, $\rho$MCTDH.\cite{raab1999,raab2000a} In this model, the dynamics in pyrazine is initiated via an instantaneous vertical excitation of the system from the electronic ground state, $S_0$, to the second excited diabatic state, $S_2\,(n,\pi^\star)$. The subsequent non-adiabatic dynamics proceed via internal conversion through a conical intersection between the electronically excited states $S_2\,(n,\pi^\star)$ and $S_1\,(\pi,\pi^\star)$. All calculations considered here were performed with the \textit{Heidelberg MCTDH package}\cite{heidelbergmctdh} in its recent version 8.5.

\subsection{The Pyrazine Model Hamiltonian}
The minimal model of the pyrazine problem is given in terms of a 2-state-4-mode vibronic coupling Hamiltonian for the electronically excited diabatic state subspace, following the notation of Ref.\cite{raab2000a}, with Hamiltonian 
\begin{equation}
\hat{H}
=
\sum^2_{i=1}
\left(E_i+\hat{H}_i\right)
\ket{S_i}
\bra{S_i}
+
\hat{V}
\biggl(
\ket{S_1}
\bra{S_2}
+
\ket{S_2}
\bra{S_1}
\biggr) \quad .
\label{eq.pyrazine_system_4d}
\end{equation}
Here $\hat{H}_i$ is the on-diagonal vibrational Hamiltonian and $\hat{V}$ the inter-state vibronic coupling. The energies $E_i$ are given as $E_1=-\Delta$ and $E_2=+\Delta$, which resemble the energy gap of $2\Delta$ between the $S_1$- and $S_2$-potential energy surfaces at the ground state equilibrium position. 
In second quantization representation, $\hat{H}_i= \hat{H}_0+\hat{V}_{c,i}=\hat{H}_0+\hat{H}^{(1)}_i+\hat{H}^{(2)}_i$, is given by
\begin{align}
\hat{H}_0
&=
\sum_{k=10a,6a,9a,1}
\hbar\omega_k
\left(
\hat{a}^\dagger_k
\hat{a}_k
+
\dfrac{1}{2}
\right),
\nonumber
\vspace{0.2em}
\\
\hat{H}^{(1)}_i
&=
\sum_{k=6a,9a,1}
\dfrac{a^{(i)}_k}{\sqrt{2}}
\left(
\hat{a}^\dagger_k
+
\hat{a}_k
\right),
\nonumber
\vspace{0.2em}
\\
\hat{H}^{(2)}_i
&=
\sum_{k,k^\prime=6a,9a,1}
\dfrac{a^{(i)}_{kk^\prime}}{2}
\left(
\hat{a}^\dagger_k
+
\hat{a}_k
\right)
\left(
\hat{a}^\dagger_{k^\prime}
+
\hat{a}_{k^\prime}
\right).
\end{align}
Here, $\hat{H}_0$ resembles the ground-state normal-mode Hamiltonian for the tuning modes, $\{v_{9a},v_{6a},v_1\}$, and the coupling mode, $v_{10a}$, with harmonic frequencies $\omega_k$. Further, $\hat{H}^{(1)}_i$ and $\hat{H}^{(2)}_i$ refer to linear and bilinear intra-state vibronic coupling terms, respectively, involving only the tuning modes $\{v_{9a},v_{6a},v_1\}$. The inter-state vibronic coupling term, $\hat{V}=\hat{V}^{(1)}+\hat{V}^{(2)}$, is composed of a linear, $\hat{V}^{(1)}$, and a quadratic interaction, $\hat{V}^{(2)}$, respectively, given by
\begin{align}
\hat{V}^{(1)}
&=
\dfrac{c_{10a}}{\sqrt{2}}
\left(
\hat{a}^\dagger_{10a}
+
\hat{a}_{10a}
\right),
\nonumber
\vspace{0.2cm}
\\
\hat{V}^{(2)}
&=
\sum_{k=6a,9a,1}
\dfrac{c_{10a,k}}{2}
\left(
\hat{a}^\dagger_{10a}
+
\hat{a}_{10a}
\right)
\left(
\hat{a}^\dagger_k
+
\hat{a}_k
\right).
\end{align}
The parameters $a^{(i)}_k,a^{(i)}_{kk^\prime},c_{10a},c_{10a,k}$ and $\omega_k$ of the linear and bilinear models are taken from Ref.\cite{raabworth1999} and are reproduced in Appendix B (\textit{cf.} Tabs.\ref{tab.parameters_mctdh_tqp_m4_lin} and \ref{tab.parameters_mctdh_tqp_m4_bilin}).
\\
Further, we consider a 20-mode harmonic oscillator bath bilinearly coupled to the linear pyrazine model\cite{worth1998} with interaction, $\hat{H}^{(SB)}_i$, and bath contribution, $\hat{H}_{B}$, given by
\begin{align}
\hat{H}^{(SB)}_i
&=
\sum_{k=1}^{20}
\dfrac{\kappa^{(i)}_k}{\sqrt{2}}
\left(
\hat{a}^\dagger_{b,k}
+
\hat{a}_{b,k}
\right),
\vspace{0.2cm}
\\
\hat{H}_B
&=
\sum_{k=1}^{20}
\hbar\omega_{b,k}
\left(
\hat{a}^\dagger_{b,k}\hat{a}_{b,k}
+
\dfrac{1}{2}
\right).
\end{align}
Here, $\{\hat{a}^\dagger_{b,k},\hat{a}_{b,k}\}$ are bosonic bath mode operators, $\hbar\omega_{b,k}$, are harmonic bath mode frequencies and, $\kappa^{(i)}_k$, are diabatic, state dependent linear bath coupling coefficients. The harmonic frequencies and bilinear coupling coefficients are considered as reported in Table I of Ref. \cite{worth1998}.
\\
In this work, we are going to study the 2-state-4-mode model system in its linear form, where we neglect the quadratic contributions, \textit{i.e.}, $\hat{H}^{(2)}_i$ and $\hat{V}^{(2)}$, a fully bilinear model taking into account second order terms and a linear 2-state-24-mode system-bath model, respectively.

\subsection{The Pyrazine Model Hamiltonian in Thermal Quasi-Particle TFD} 
In line with arguments given in Sec.(\ref{eq.nonadiabat_tqp_tfd}), the TFD representation of the pyrazine vibronic coupling Hamiltonian can be written as
\begin{equation}
\bar{H}
=
\sum^2_{i=1}
\left(E_i+\hat{H}_i-\tilde{H}_0\right)
\ket{S_i}
\bra{S_i}
+
\hat{V}
\biggl(
\ket{S_1}
\bra{S_2}
+
\ket{S_2}
\bra{S_1}
\biggr),
\label{eq.pyrazine_system_4d_tfd}
\end{equation}
where we only treat the harmonic vibrational-mode contribution within TFD, leading to
\begin{equation}
\bar{H}_0
= 
\hat{H}_0
-
\tilde{H}_0
=
\sum_{k=10a,6a,9a,1}
\hbar\omega_i
\left(
\hat{a}^\dagger_i
\hat{a}_i
-
\tilde{a}^\dagger_i
\tilde{a}_i
\right) \quad .
\end{equation}
A formally identical expression is obtained for the 20-mode harmonic bath Hamiltonian, \textit{i.e.}, $\bar{H}_B=\hat{H}_B-\tilde{H}_B$. The thermal quasi-particle representation of $\bar{H}$ is subsequently obtained via the inverse TBT relations, \textit{cf.}, Eqs.\eqref{eq.inverse_tbt_phys} and \eqref{eq.inverse_tbt_aux}, and can be written in the form
\begin{equation}
\bar{H}_\beta
=
\sum^2_{i=1}
\left(
E_i
+
\bar{H}^{(0)}_\beta
+
H^{(1)}_{\beta,i}
+
H^{(2)}_{\beta,i}
\right)
\ket{S_i}
\bra{S_i}
+
V_\beta
\biggl(
\ket{S_1}
\bra{S_2}
+
\ket{S_2}
\bra{S_1}
\biggr) \quad .
\label{eq.pyrazine_system_4d_tqp}
\end{equation}
For the extended 2-state-24-mode system-bath Hamiltonian, one augments the system Hamiltonian $\bar{H}_\beta$ as
\begin{equation}
\bar{H}_\beta^{SB}
=
\bar{H}_\beta
+
\sum^2_{i=1}
\left(
H^{(SB)}_{\beta,i}
+
\bar{H}^{(B)}_\beta
\right)
\ket{S_i}
\bra{S_i}
\quad .
\label{eq.pyrazine_system_24d_tqp}
\end{equation}
Explicit expressions for $\bar{H}^{(0)}_\beta,H^{(1)}_{\beta,i},H^{(2)}_{\beta,i},V_\beta$ and $\bar{H}^{(B)}_\beta,H^{(SB)}_{\beta,i}$ are given in Appendix C. 

\subsection{Observables at Finite Temperature}
We calculate observables and linear absorption spectra from the time-evolution of the vibronic thermofield state $\ket{\Psi_\beta(t)}$. As initial state, we consider
\begin{equation}
\ket{\Psi_\beta(t_0)}
=
\ket{S_2}\ket{\underline{0}_\beta}
=
\left(
\hat{\mu}\ket{S_0}
\right)
\ket{\underline{0}_\beta},
\end{equation}
where we assume an instantaneous excitation from the electronic ground state mediated by the electronic dipole operator $\hat{\mu}=\mu_{20}\left(\ket{S_0}\bra{S_2}+\ket{S_2}\bra{S_0}\right)$ (we set $\mu_{20}=1$ in what follows). Further, $\ket{\underline{0}_\beta}=\ket{0^{v10a}_\beta}\ket{0^{v6a}_\beta}\ket{0^{v9a}_\beta}\ket{0^{v1}_\beta}$, is the vibrational thermal vacuum state of the harmonic 4-mode-model and, $\ket{\underline{0}^{(B)}_\beta}=\prod^{20}_{k=1}\ket{0^{(b),k}_\beta}$, the harmonic-bath thermal vacuum state, respectively. For the linear 2-state-24-mode model, which comprises 48 vibrational DoFs in the TFD framework, the initial state accordingly reads, $\ket{\Psi_\beta(t_0)}=\ket{S_2}\ket{\underline{0}_\beta}\ket{\underline{0}^{(B)}_\beta}$.
\\
Further, time-evolution of the electronic DoF is studied by means of electronic diabatic populations $P^{(S_1)}_\beta(t),P^{(S_2)}_\beta(t)$, given as
\begin{equation}
P^{(S_i)}_\beta(t)
=
\braket{
\Psi_\beta(t)
\vert
\biggl(
\ket{S_i}\bra{S_i}
\biggr)
\vert
\Psi_\beta(t)}
\hspace{0.5cm}
i=1,2,
\label{eq.tfd_electronic_pop}
\end{equation}
where the connection to the quantum statistical expression, $P^{(S_i)}_\beta(t)=\mathrm{tr}\{\hat{\rho}(t)\ket{S_i}\bra{S_i}\}$, is explicitly given in {Appendix D}. Moreover, vibrational dynamics of the coupling mode, $v_{10a}$, and the tuning modes, $\{v_{9a},v_{1},v_{6a}\}$, respectively, are studied by their mean occupation numbers $\braket{\hat{n}_k}_\beta(t)$, given by Eq.\eqref{eq.tfd_vib_pop} in {Appendix D}.
Every vibrational mode is initially in a thermal equilibrium state.

Finally, we calculate linear absorption spectra at finite temperature via
\begin{equation}
\sigma_\beta(\omega)
\propto
\mathrm{Im}
\displaystyle\int^\infty_0\,
C_\beta(t)\,
e^{\text{i}\omega\,t}\,
\mathrm{d}t \quad ,
\label{eq.tfd_abs_spec}
\end{equation}
where $C_\beta(t)$ is the thermal autocorrelation function given by
\begin{equation}
C_\beta(t)
=
\braket{
\Psi_\beta
\vert
e^{-\text{i}\bar{H}_\beta\,t/\hbar}
\vert
\Psi_\beta}
=
\mathrm{tr}
\left\{
\biggl(
e^{\text{i}\hat{H}\,t/\hbar}
\hat{\mu}\,
e^{-\text{i}\hat{H}\,t/\hbar}
\hat{\mu}\,
\biggr)\,
\hat{\rho}(t_0)
\right\},
\label{eq.tfd_acf}
\end{equation}
with $\ket{\Psi_\beta(t)}=e^{-\text{i}\bar{H}_\beta\,t/\hbar}\ket{\Psi_\beta}$ and $\hat{\rho}(t_0)=\hat{\rho}^0_\beta\ket{S_0}\bra{S_0}$. A detailed derivation of Eq. \eqref{eq.tfd_acf}, which constitutes a special case of very recently published general results by Gelin and Borrelli\cite{gelin2021}, is presented in Appendix E.

\section{Results and Discussion}
\label{sec.results}
We discuss numerical results for the pyrazine model in its linear and bilinear 2-state-4-mode variants and the linear 2-state-24-mode system-bath version. We first give a detailed examination of the multilayer expansion of the thermofield states and compare the computational performance for selected multilayer trees based on CPU time. Afterwards, we discuss temperature effects on internal conversion dynamics in the respective pyrazine models as well as corresponding linear absorption spectra evaluated for different temperatures. Finally, we will consider the impact of a bilinearly coupled harmonic bath on the linear 2-state-4-mode system at finite temperature.

\subsection{Multilayer Expansion of Thermal Quasi-Particle States}
\begin{figure}[hbt]
\includegraphics[scale=1.0]{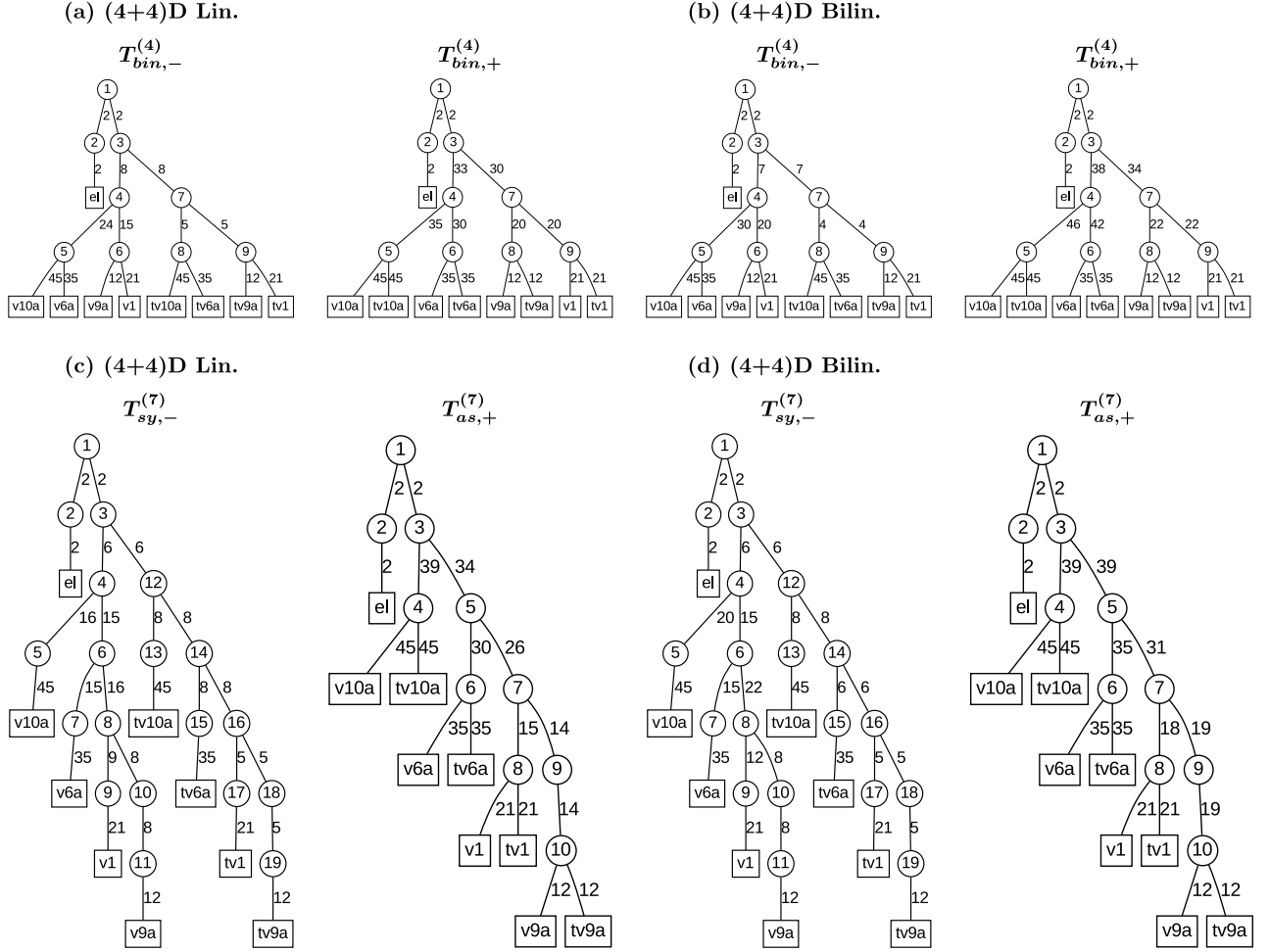}
\caption{Multilayer trees for thermofield states of linear (lin.) and bilinear (bilin.) (4+4)D-pyrazine models with number of tSPFs given next to edges (at bottom layer, numbers correspond to primitive TQP number states). Physical primitive modes are specified as $\{v10a,\,v6a,\,v9a,\,v1\}$ and auxiliary primitive modes as $\{tv10a,\,tv6a,\,tv9a,\,tv1\}$. Top row: Symmetric binary 4-layer trees, $T^{(4)}_{bin}$, for (a) linear and (b) bilinear (4+4)D-model with low-temperature ($T^{(4)}_{bin,-}$) and high-temperature ($T^{(4)}_{bin,+}$) mode combination schemes. Bottom row: Symmetric, $T^{(7)}_{sy,-}$, and asymmetric, $T^{(7)}_{as,+}$, 7-layer trees for the (c) linear and (d) bilinear (4+4)D-mode model with low-temperature (index "$-$") and high-temperature (index "$+$") mode combination schemes. }
\label{fig.multilayer_tqp_tree_fig}
\end{figure}

We discuss the multilayer expansions of TQP thermofield states for 2-state-4-mode, abbreviated as (4+4)D, models. A conventional graphical representation of the multilayer expansion is given in terms of loop-free undirected graphs\cite{manthe2008} denoted as ML-trees (\textit{cf.}, Fig.(\ref{fig.multilayer_tqp_tree_fig}) for examples). 
A circle in a ML-tree corresponds to a set of tensorial coefficients of rank equals the number of connected edges and numbers next to the edges indicate the number of SPFs; primitive nodes are indicated by squares, respectively. We compare different multilayer trees for thermofield states, as shown in Fig.(\ref{fig.multilayer_tqp_tree_fig}), with respect to numerical performance (CPU time) at different temperatures for the linear and the bilinear (4+4)D-models.

In the first row of Fig.(\ref{fig.multilayer_tqp_tree_fig}), symmetric binary 4-layer trees, $T^{(4)}_{bin}$, are shown with low-temperature, $T^{(4)}_{bin,-}$, and high-temperature, $T^{(4)}_{bin,+}$, mode combination schemes for both (4+4)D models, respectively. For the low-temperature scenario, we employed combined modes $(v_{10a}, v_{6a}), (v_{9a}, v_1), (tv_{10a}, tv_{6a})$ and $(tv_1,tv_{9a})$ following Ref.\cite{raab1999}, where physical and auxiliary DoFs were separated. In the high-temperature regime, we considered mixed combined modes of the form $(v_{10a}, tv_{10a}), (v_{6a}, tv_{6a}), (v_1, tv_1)$ and $(v_{9a},tv_{9a})$, respectively, where we paired a physical mode with its respective auxiliary partner.

In the second row of Fig.(\ref{fig.multilayer_tqp_tree_fig}), we present a symmetric binary 7-layer tree, $T^{(7)}_{sy}$, and a corresponding asymmetric version, $T^{(7)}_{as}$, where the individual branches are truncated at different layers. We employ the same temperature-dependent mode-combination scheme as for the 4-layer trees, which leads to a low-temperature tree, $T^{(7)}_{sy,-}$, and a high-temperature version, $T^{(7)}_{as,+}$, respectively. Further, the structures of all ML-trees are chosen identical for both the linear and the bilinear (4+4)D-models, however, they differ in the number of tSPFs, respectively.
\\
Turning to the numerical performance, we discuss CPU times for converged propagation runs obtained with the presented ML-trees for the linear (\textit{cf.} Tab.\ref{tab.pyr_m4_linear_cpu}) and bilinear (\textit{cf.} Tab.\ref{tab.pyr_m4_bilinear_cpu}) (4+4)D-model with propagation time $t_f=150\,\text{fs}$ for different temperatures. A calculation is identified as converged, if the highest natural population satisfies $\leq1.0\times10^{-3}$. All multilayer calculations are compared to results from a standard 2-layer MCTDH-TQP expansion scheme of the thermofield state, indicated by the symbol $T^{(2)}_0$ (see Appendix F for numerical details). 
\begin{table}[hbt]
    \caption{CPU time (h:m) for Linear and bilinear (4+4)D-model Hamiltonians of Pyrazine with different ML-tree topologies and propagation time $t_f=150\,\text{fs}$ (Intel(R) Xeon(R) CPU E5-2650 v2 @ 2.60GHz, 126 GB RAM) for a pyrazine (4+4)D model at different temperatures. 
    }
    \vspace{0.5cm}
    \centering
    Linear (4+4)D-Model
    \vspace{0.2cm}\\
    \begin{tabular}{c c c c c c c c c c}
       \hline
        Temp. \hspace{1.5em}          & $T^{(4)}_{bin,-}$  && $T^{(4)}_{bin,+}$  &&  $T^{(7)}_{sy,-}$  &&  $T^{(7)}_{as,+}$  &&  $T^{(2)}_0$  
              \vspace{0.3em}\\ 
       \hline\hline\\
       $1\,\text{K}$\hspace{1.5em}    & 0:06 && 0:08 && 0:02 && 0:05 && 0:03
       \vspace{0.2em}\\ 
       $100\,\text{K}$\hspace{1.5em}  & 0:08 && 0:17 && 0:04 && 0:08 && 0:04
       \vspace{0.2em}\\
       $300\,\text{K}$\hspace{1.5em}  & 3:54 && 0:29 && 0:24 && 1:02 && 0:05
       \vspace{0.2em}\\
       $500\,\text{K}$\hspace{1.5em}  & 3:07 && 0:52 && 3:09 && 0:53 && 0:29
       \vspace{0.3em}\\
       \hline
    \end{tabular}
    \label{tab.pyr_m4_linear_cpu}
\end{table}
For both models and all ML-trees, we observe an increase in CPU time with increasing temperature, which is directly related to a temperature-dependent increase of interactions in the TQP Hamiltonian (\textit{cf.} Eq.\eqref{eq.pyrazine_system_4d_tqp}). Further, the low- and high-temperature (here, $T=500\,\mathrm{K}$) expansions perform significantly better in their respective regimes and for high temperatures, a CPU time reduction of factors 2-4 depending on the ML-tree is observed.
\begin{table}[hbt]
    \caption{CPU time (h:m) for Linear and bilinear (4+4)D-model Hamiltonians of Pyrazine with different ML-tree topologies and propagation time $t_f=150\,\text{fs}$ (Intel(R) Xeon(R) CPU E5-2650 v2 @ 2.60GHz, 126 GB RAM) for a pyrazine (4+4)D model at different temperatures. 
    }
    \vspace{0.5cm}
    \centering
    Bilinear  (4+4)D-Model
    \vspace{0.2cm}\\
    \begin{tabular}{c c c c c c c c c c }
       \hline
        Temp. \hspace{1.5em}          & $T^{(4)}_{bin,-}$  &&  $T^{(4)}_{bin,+}$  &&  $T^{(7)}_{sy,-}$  &&  $T^{(7)}_{as,+}$  && $T^{(2)}_0$ \vspace{0.3em}\\ 
       \hline\hline\\
       $1\,\text{K}$\hspace{1.5em}   & 0:19 && 0:21 && 0:05 && 0:17 && 0:15
       \vspace{0.2em}\\ 
       $100\,\text{K}$\hspace{1.5em}  & 0:38 && 1:06 && 0:10 && 0:39 && 0:37 
       \vspace{0.2em}\\
       $300\,\text{K}$\hspace{1.5em}  & 3:29 && 2:12 && 1:25 && 1:29 && 3:01
       \vspace{0.2em}\\
       $500\,\text{K}$\hspace{1.5em}  & 7:16 && 4:17 && 8:41 && 2:44 && 9:36
       \vspace{0.3em}\\
       \hline
    \end{tabular}
    \label{tab.pyr_m4_bilinear_cpu}
\end{table}
Moreover, we find the 7-layer trees, $T^{(7)}_{sy,-}$ and $T^{(7)}_{as,+}$, to outperform their symmetric binary 4-layer counterparts, $T^{(4)}_{bin,-}$ and $T^{(4)}_{bin,+}$, respectively. For temperatures up to $T=300\,\text{K}$, the shortest CPU times for both (4+4)D models are observed for $T^{(7)}_{sy,-}$, while at $T=500\,\text{K}$, the performance of $T^{(7)}_{as,+}$ takes over. In this context, we observe a more compact wave function representation for the high-temperature mode combination scheme and a more efficient memory usage in the corresponding propagation runs. However, the numerical details of the high-temperature scheme explaining its computational advantage are not yet fully resolved. We finally remark, that for the linear model a standard MCTDH expansion is numerically sufficient and gives the shortest CPU times for all temperatures considered here, whereas the multilayer expansion is to be preferred for the bilinear model with more complex interactions. 
\\
We close by commenting on the relation of the MCTDH-TQP approach to the stochastic MCTDH approaches and the $\rho$MCTDH method. The former, \textit{e.g.}, in Refs.\cite{manthe2001,nest2007}, require several system-dependent realizations of an imaginary-time propagation of a randomly created initial state followed by a real-time propagation to properly converge a time-dependent ensemble average. For many DoFs, the use of the ML-MCTDH method becomes mandatory and this approach becomes numerically demanding. However, at the moment it remains an open question which approach allows for numerically faster results with respect to a certain Hamiltonian and for certain properties, which may be determined by "frequent" or "infrequent" events.
Further, the density operator formulation of MCTDH, \textit{i.e.}, $\rho$MCTDH, has shown to be numerically efficient for small systems\cite{raab1999}, however, is at the moment restricted to a small number of DoFs. The presented MCTDH-TQP approach requires only a single realization and is able to treat a large number of DoFs due to the advantages of the ML-MCTDH expansion for the calculation of thermal ensemble averages. In conclusion, we find that temperature effects can be efficiently included in the multilayer expansion of thermofield states and a proper comparison with the stochastic and density operator MCTDH approaches is desirable to reveal important details about their preferred fields of application.

\subsection{Thermal Effects on Internal Conversion in (4+4)D-Pyrazine Models}
We study thermal effects on electronic and vibrational dynamics in linear and bilinear (4+4)D-pyrazine models employing the MCTDH-TQP approach. Our findings are compared to results based on the $\rho$MCTDH approach, which have been reported by Raab \textit{et al.}\cite{raab1999} and reproduced here via the \textit{Heidelberg MCTDH package}, Version 8.3\,\cite{heidelbergmctdh}.
\begin{figure}[hbt]
\includegraphics[scale=1.0]{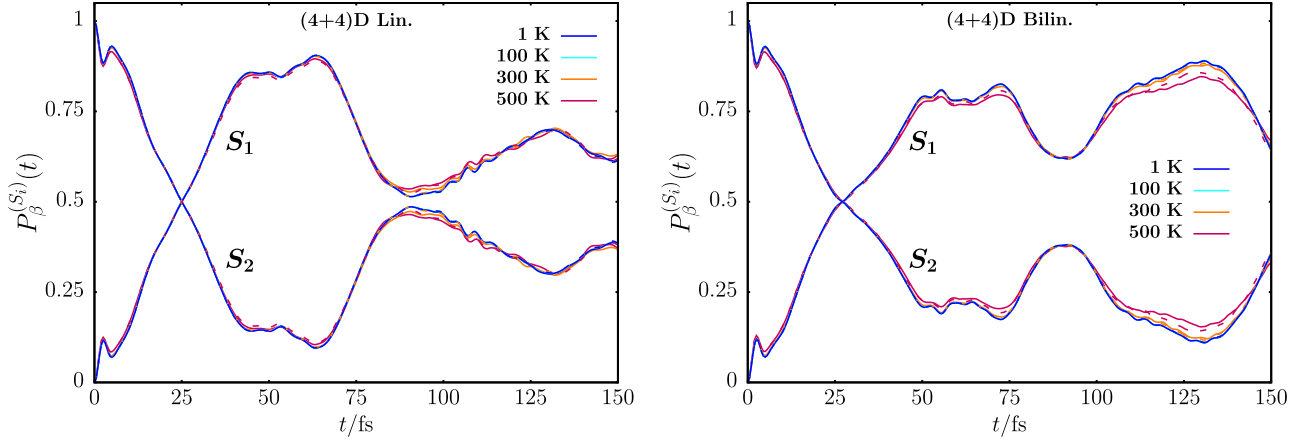}
\caption{Time evolution of diabatic electronic populations $P^{(S_1)}_\beta(t)$ and $P^{(S_2)}_\beta(t)$ for electronic states $\ket{S_1}$ and $\ket{S_2}$ for the linear (left) and bilinear (right) (4+4)D pyrazine model and $\rho$MCTDH results (dashed lines; same color code as MCTDH-TQP) reproduced from Ref.\cite{raab1999} for temperatures $T=1\,\text{K},\,100\,\text{K},300\,\text{K}$ and $500\,\text{K}$}
\label{fig.epop_temp_fig}
\end{figure}

\begin{figure}[hbt]
\includegraphics[scale=1.0]{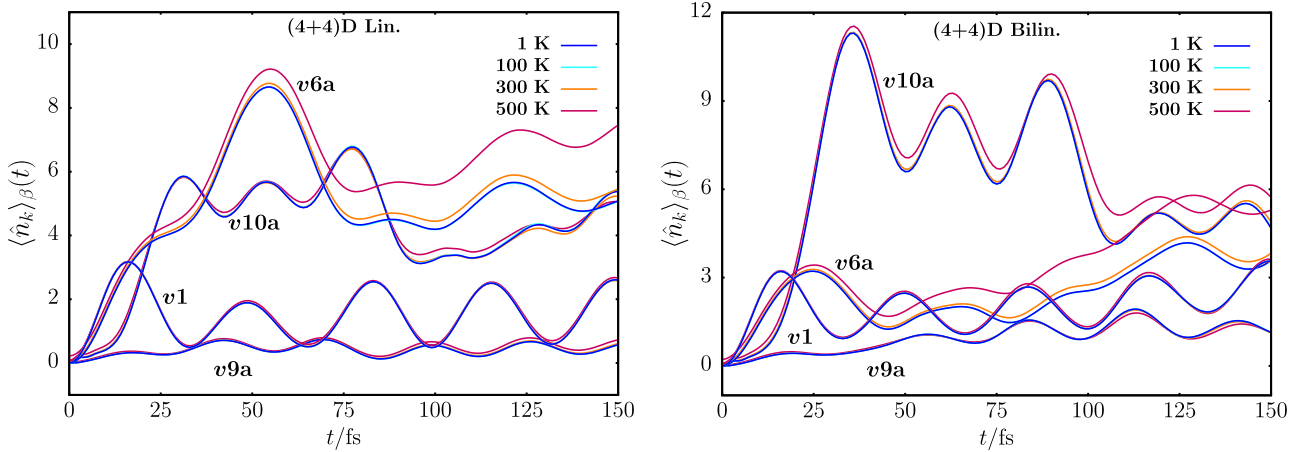}
\caption{Time-evolution of the state averaged thermal mean occupation numbers $\braket{\hat{n}_k}_\beta(t)$ of the coupling mode $v10a$ and the tuning modes $\{v6a,v9a,v1\}$  for the linear (left) and bilinear (right) (4+4)D pyrazine model for temperatures $T=1\,\text{K},\,100\,\text{K},300\,\text{K}$ and $500\,\text{K}$.}
\label{fig.modes_temp_fig}
\end{figure}

In Fig.(\ref{fig.epop_temp_fig}), the time evolution of diabatic electronic populations, $P^{(S_1)}_\beta(t)$ and $P^{(S_2)}_\beta(t)$, is shown for the linear (left) and the bilinear model (right). The results are obtained via the corresponding thermofield state expansions, which require the shortest propagation time according to Tabs.\ref{tab.pyr_m4_linear_cpu} and \ref{tab.pyr_m4_bilinear_cpu}.
\\
For the linear model, the dynamics is characterized by a fast initial decay and a pronounced recurrence around $100\,\text{fs}$. At finite temperature, we observe slightly faster relaxation at small times and some small temperature dependent damping effects on the recurrence. For the bilinear model, we also observe slightly faster initial relaxation and temperature dependent damping effects of recurrences at around $60\,\text{fs}$ and $120\,\text{fs}$, respectively. Due to stronger interaction contributions in the TQP-Hamiltonian, the temperature effects are more pronounced in the bilinear (4+4)D-model. In comparison to $\rho$MCTDH-results, we find diabatic electronic populations behaving similar at short times and showing a slightly stronger population transfer at later times. A possible reason for this small deviation might result from the different representation of the Hamiltonian in Ref.\cite{raab1999} compared to the second quantization representation employed here.
\\
Turning to the vibrational modes of the (4+4)D-models (\textit{cf.} Fig.(\ref{fig.modes_temp_fig})), we consider the time evolution of thermal mean occupation numbers, $\braket{\hat{n}_k}_\beta(t)$, at different temperatures. 
In general, we can distinguish the dynamics of high-frequency modes, $(v9a,v1)$, and low-frequency modes, $(v6a,v10a)$, respectively. The dynamics of the high-frequency modes is characterized by low amplitude Rabi-type oscillations and minor shifts to slightly higher mean occupation numbers. In contrast, the low frequency modes show some significant thermal excitations. In particular, for the tuning mode, $v6a$, we observe relatively strong thermal excitation in the high-temperature regime for both models. Additionally, also the coupling mode, $v10a$, shows more pronounced thermal effects in the bilinear model, which can be traced back to its coupling to the tuning modes. 

\subsection{Thermal Effects on Spectra}
\begin{figure}[hbt]
\includegraphics[scale=1.0]{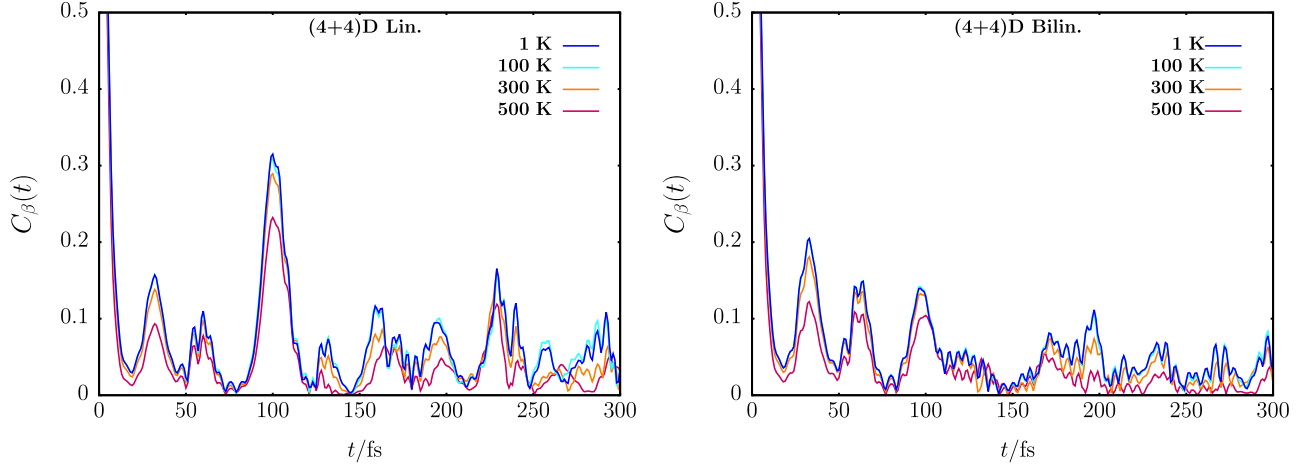}
\caption{Time-evolution of the TFD autocorrelation function $C_\beta(t)$ for the linear (left) and bilinear (right) (4+4)D pyrazine model for temperatures $T=1\,\text{K},\,100\,\text{K},300\,\text{K}$ and $500\,\text{K}$.}
\label{fig.auto_temp_fig}
\end{figure}

\begin{figure}[hbt]
\includegraphics[scale=1.0]{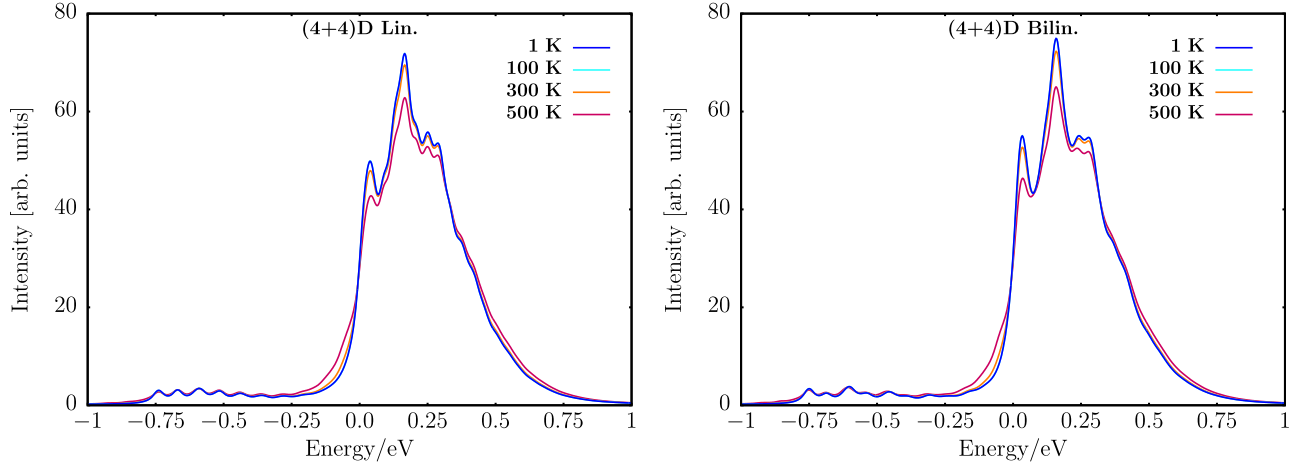}
\caption{Temperature-dependent linear absorption spectra $\sigma_\beta(\omega)$ calculated from the TFD autocorrelation function $C_\beta(t)$ for the linear (left) and bilinear (right) (4+4)D pyrazine model for temperatures $T=1\,\text{K},\,100\,\text{K},300\,\text{K}$ and $500\,\text{K}$.}
\label{fig.spec_temp_fig}
\end{figure}

Here, we turn to the time-evolution of the system from the perspective of the thermal autocorrelation function, $C_\beta(t)$, (\textit{cf.} Fig.(\ref{fig.auto_temp_fig})), which allows us to subsequently access linear absorption spectra (\textit{cf.} Fig.(\ref{fig.spec_temp_fig})).
For both (4+4)D-models, the time-evolution of $C_\beta(t)$ is determined by a rapid initial decay in the first $20\,\text{fs}$, followed by a series of recurrences. The bilinear model is characterized by a collection of damped recurrences, whereas in the linear model a particularly pronounced recurrence is observed at around $100\,\text{fs}$. Finite temperature effects manifest as a slightly faster initial decay and a damping of the recurrences for both systems. From the thermal autocorrelation function, we calculate the corresponding linear absorption spectra, $\sigma_\beta(\omega)$, for the linear and bilinear 2-state-4-mode model on the left- and right-hand side of Fig.(\ref{fig.spec_temp_fig}), respectively.
The presentation is based on Ref.\cite{worth1998}, where the energy axis is chosen such, that the energy-zero resembles half the energy difference between the diabatic states $S_1$ and $S_2$ at the ground state equilibrium geometry. Additionally, the thermal autocorrelation function has been exponentially damped via $\exp(-t/\tau)$ with $\tau=30\,\text{fs}$ in Eq.\eqref{eq.tfd_abs_spec} to phenomenologically account for the remaining twenty modes of pyrazine compared to the 4-mode models\cite{raab1999}. For both models, we observe a broadening accompanied by intensity reduction of the peak at positive energies with increasing temperature. For the low intensity peaks at negative energy, thermal effects are significantly weaker.
\subsection{Thermal Effects with Harmonic-Oscillator Bath}
Finally, we consider the impact of a linearly coupled harmonic oscillator bath on the linear 2-state-4-mode model at finite temperature. The resulting extended linear 2-state-24-mode system-bath model, as given by the Hamiltonian in Eq.\eqref{eq.pyrazine_system_24d_tqp}, exceeds the actual capabilities of the $\rho$MCTDH approach and comprises 48 vibrational DoFs in total, 24 physical and 24 auxiliary ones, \textit{i.e.}, a (24+24)D model, in the TFD framework. 

\begin{figure}[hbt]
\includegraphics[scale=1.0]{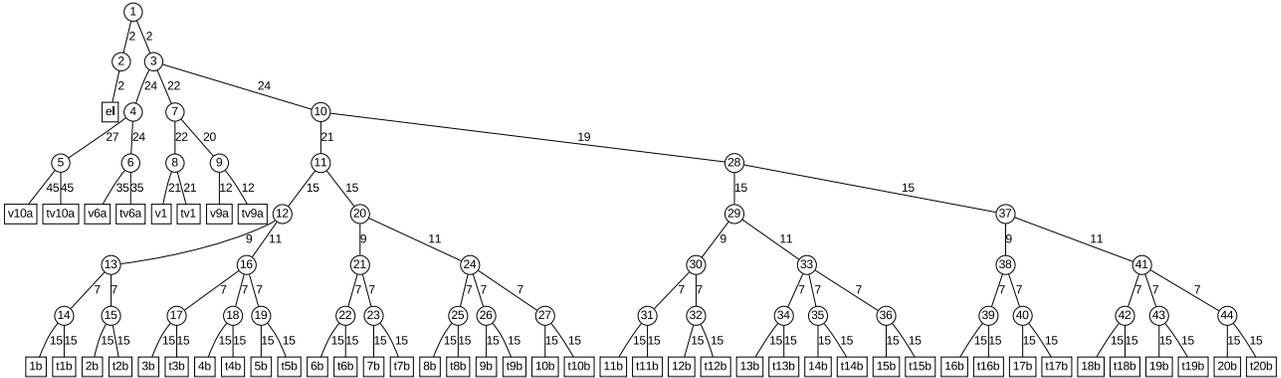}
\caption{Exemplary ML-tree for a vibronic thermofield state of the linear 2-state-24-mode pyrazine model.}
\label{fig.ml_tree_sys_bath}
\end{figure}

The multilayer expansion of the vibronic thermofield state follows the binary structure of $T^{(4)}_{bin}$ (\textit{cf.} Fig.\ref{fig.multilayer_tqp_tree_fig}) for the system and for the bath, we add an additional 5-layer sub-tree at node 3 (\textit{cf.} Fig.\ref{fig.ml_tree_sys_bath}).

The mode combination procedure follows the low- and high-temperature scheme discussed above, where the high-temperature scheme was applied for both $300\,\mathrm{K}$ and $500\,\mathrm{K}$. 
\begin{figure}[hbt]
\includegraphics[scale=1.0]{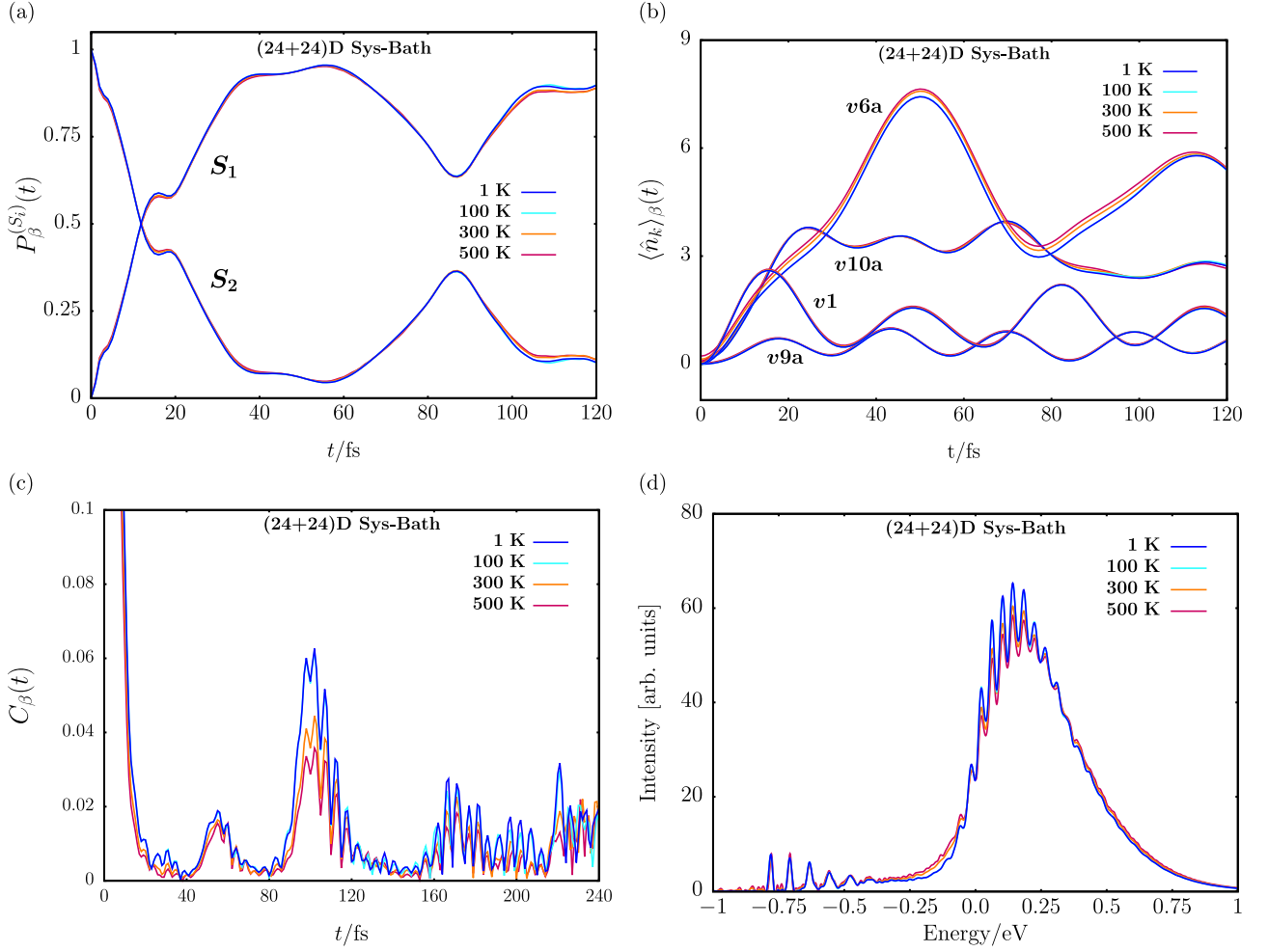}
\caption{(a) Time-evolution of diabatic electronic populations $P^{(S_i)}_\beta(t)$; (b) time-evolution of the state-averaged thermal mean occupation numbers $\braket{\hat{n}_k}_\beta(t)$ of system modes; (c) time-evolution of the TFD autocorrelation function $C_\beta(t)$; (d) temperature-dependent linear absorption spectra, $\sigma_\beta(\omega)$. All for the linear (24+24)D system-bath pyrazine model for temperatures $T=1\,\text{K},\,100\,\text{K},300\,\text{K}$ and $500\,\text{K}$.}
\label{fig.sys_bath_temp_fig}
\end{figure}

The diabatic populations (\textit{cf.} Fig.\ref{fig.sys_bath_temp_fig}(a)) show a more complete decay of the $S_2$ state due to the presence of the bath with a small recurrence around $50\,\mathrm{fs}$ and a large one around $90\,\mathrm{fs}$. The influence of thermal energy on the diabatic populations is rather small for the temperatures studied here as already observed in the case of the linear (4+4)D model, which can be traced back to the relative high frequencies of the system and bath modes. More pronounced but yet weak thermal effects are observed for the coupling mode, $v_{6a}$, (\textit{cf.} Fig.\ref{fig.sys_bath_temp_fig}(b)) and the TFD autocorrelation function $C_\beta(t)$ (\textit{cf.} Fig.\ref{fig.sys_bath_temp_fig}(c)): The coupling mode is thermally excited, whereas the autocorrelation function exhibits a thermal damping, which directly manifests as thermal broadening in the corresponding linear absorption spectrum (\textit{cf.} Fig.\ref{fig.sys_bath_temp_fig}(d)). We note, that the linear absorption spectrum shown here is only broadened due to the explicit presence of the bath modes, \textit{i.e.}, no artificial broadening has been considered opposed to the (4+4)D models discussed above, which results in a richer structure compared to Fig.\ref{fig.spec_temp_fig}.

\section{Summary and Conclusions}
\label{sec.conclusion}
We introduced a thermofield-based formulation of the multilayer multiconfigurational time-dependent Hartree (ML-MCTDH) method for the treatment of non-adiabatic quantum dynamics at finite temperature. Our work is based on the thermal quasi-particle (TQP) representation of symmetric thermofield dynamics (TFD), which provides a formulation of quantum statistical mechanics in the language of many-body theory. We introduced the ML-MCTDH approach for thermofield states by exploiting the formal equivalence of thermal quasi-particle TFD at fixed finite temperature and bosonic many-body theory at zero temperature with doubled number of degrees of freedom. This equivalence allows for a transfer of bosonic many-body MCTDH, \textit{i.e}., MCTDH-SQR, to the  thermal quasi-particule TFD framework. In particular, the thermal quasi-particle representation of TFD provides an appealing form of the multi-mode thermal vacuum state in terms of a Hartree product of single-mode thermal vacua, which constitutes an ideal initial state for the MCTDH method. From a practical point of view, the method presented here can be directly applied via the ML-MCTDH-SQR method as implemented in the \textit{Heidelberg MCTDH package}\cite{heidelbergmctdh}.

We applied our ansatz, abbreviated as MCTDH-TQP to emphasize the temperature dependence, to the well-studied 2-state-4-mode vibronic coupling model of pyrazine and its 2-state-24-mode extension including an additional bilinearly coupled harmonic bath. From a computational point of view, the effect of temperature, which manifests as increased interactions in the TQP thermofield Hamiltonian, can be beneficially accounted for in the multilayer expansion of thermofield states by properly combining physical and auxiliary DoFs. For the multilayer trees used in this work, it turned out to be computationally advantageous to fully separate physical and auxiliary DoF at low temperatures and combine two similar physical and auxiliary modes at high temperatures. 

For the 2-state-4-mode pyrazine models, we studied finite temperature effects on the time-evolution of electronic populations and the state averaged thermal mean occupation numbers of vibrational modes. With increasing temperature, the electronic populations decay slightly faster and exhibit weak temperature depend dampings of recurrences, whereas mean occupation numbers of low frequency vibrational modes increase at elevated temperature and long propagation times. Additionally, we studied the time evolution of thermal autocorrelation functions, which can be related to correlation functions in quantum statistical mechanics and subsequently allowed for the calculation of linear absorption spectra at different temperatures. We find our results to compare well to existing results for pyrazine at elevated temperature based on the $\rho$MCTDH ansatz for density matrices. In order to go beyond the actual capabilities of the $\rho$MCTDH ansatz, we additionally considered a linear 2-state-24-mode system-bath model accessible via the ML-MCTDH-TQP approach. Here, we found small but non-negligible thermal effects for the temperature regimes studied here.

A very promising route to future applications of the ML-MCTDH-TQP approach is provided by a recently formulated and refined dynamical spawning of single particle functions\cite{tapia2017,tapia2021}, which allows for fully automatized convergence of MCTDH calculations with respect to SPFs. This significantly simplifies both high-dimensional ML-MCTDH-TQP calculations and the optimization of ML-trees, which is highly desirable facing the large number of DoFs in TFD. {Further, a comparison of the numerical performance of MCTDH-TQP with respect to stochastic and density operator approaches in the MCTDH framework as well as the numerical details of the high-temperature mode combination scheme is desirable. Finally, the discussion of formal connections between the $\rho$MCTDH and MCTDH-TQP methods presumably enhances both the future optimization and further understanding of these methods.}

\section*{Acknowledgments}
E.W. Fischer gratefully acknowledges fruitful discussions with Prof. Dr. Oriol Vendrell (Heidelberg) and the kind hospitality of Prof. Vendrell's group. E.W. Fischer gratefully acknowledges support on MCTDH-SQR and helpful feedback on this manuscript by Prof. Dr. Hans-Dieter Meyer (Heidelberg). E.W. Fischer gratefully acknowledges helpful feedback on this manuscript and fruitful discussions with Dr. David Picconi (Potsdam). The authors thank the Deutsche Forschungsgemeinschaft (DFG) for financial support through project Sa 547/9. E.W. Fischer acknowledges support by the International Max Planck Research School for Elementary Processes in Physical Chemistry.

\section*{Conflict of Interest}
The authors have no conflicts to disclose.
\section*{Data Availability Statement}
The data that support the findings of this study are available from the corresponding author upon reasonable request.
\newpage

\section*{Appendix A: Thermal Number States and Operators}
{The bosonic creation/annihilation operators for physical, $\{\hat{a}^\dagger_k,\hat{a}_k\}$, and auxiliary vibrational DoF, $\{\tilde{a}^\dagger_k,\tilde{a}_k\}$, respectively, satisfy canonical bosonic commutation relations
\begin{equation}
\left[\hat{a}_k,\hat{a}^\dagger_{k^\prime}\right]
=
\left[\tilde{a}_k,\tilde{a}^\dagger_{k^\prime}\right]
=
\delta_{kk^\prime},
\end{equation}
whereas all remaining commutators vanish identically. For the $k^{\text{th}}$ mode, the two-mode vacuum state $\ket{0_k,\tilde{0}_k}$ is characterized by the relations $\hat{a}_k\ket{0_k,\tilde{0}_k}=\tilde{a}_k\ket{0_k,\tilde{0}_k}=0$. From the multi-mode vacuum state $\ket{\underline{0},\underline{\tilde{0}}}$, orthonormal thermal number states can directly be constructed via $\hat{a}^\dagger_k,\tilde{a}^\dagger_k$ as
\begin{equation}
\ket{n_1,\dots,n_f,\tilde{m}_1,\dots,\tilde{m}_f}
=
\prod^{f}_{k,k^\prime=1}
\dfrac{\left(
\hat{a}^\dagger_k
\right)^{n_k}}{\sqrt{n_k!}}
\dfrac{\left(
\tilde{a}^\dagger_{k^\prime}
\right)^{m_{k^\prime}}}{\sqrt{m_{k^\prime}!}}
\ket{\underline{0},\tilde{\underline{0}}},
\label{eq.thermal_numberstates_definition}
\end{equation}
which span with $\ket{\underline{0},\underline{\tilde{0}}}$ the thermal Fock space $\mathcal{H}_\beta$ and therefore provide a representation of a general vibrational thermofield state $\ket{\psi_\beta(t)}$ on $\mathcal{H}_\beta=\mathcal{H}\otimes\tilde{\mathcal{H}}$. 
\\
The TQP operators $\hat{b}_k$ and $\tilde{b}_k$ are obtained from $\hat{a}^\dagger_k,\hat{a}_k$ and $\tilde{a}^\dagger_k,\tilde{a}_k$ via the thermal Bogoliubov transformation (TBT)
\begin{align}
\hat{b}_k
&=
\cosh\theta_k(\beta)\,
\hat{a}_k
-
\sinh\theta_k(\beta)\,
\tilde{a}^\dagger_k,
\label{eq.physical_tqp_operator}
\vspace{0.2em}
\\
\tilde{b}_k
&=
\cosh\theta_k(\beta)\,
\tilde{a}_k
-
\sinh\theta_k(\beta)\,
\hat{a}^\dagger_k.
\label{eq.auxiliary_tqp_operator}
\end{align}
and satisfy the canonical commutation relations $[\hat{b}_k,\hat{b}^\dagger_{k^\prime}]=[\tilde{b}_k,\tilde{b}^\dagger_{k^\prime}]=\delta_{kk^\prime}$, whereas all remaining commutators vanish identically. Corresponding creation operators, $\hat{b}^\dagger_k$ and $\tilde{b}^\dagger_k$, follow directly as hermitian conjugates of Eqs.\eqref{eq.physical_tqp_operator} and \eqref{eq.auxiliary_tqp_operator}, respectively. For bosonic modes, the thermal mixing angles $\theta_k(\beta)$ are defined as\cite{takahashi1996} 
\begin{equation}
\theta_k(\beta)
=
\mathrm{arctanh}\left(e^{-\beta\hbar\omega_k/2}\right)
\label{eq.harmonic_tqp_mixing_angle}
\end{equation}
with harmonic frequencies $\omega_k$. The latter follows from the identity\cite{khanna2009}
\begin{equation}
\hat{a}_k\ket{0^{(k)}_\beta}
=
e^{-\beta\omega_k/2}\,
\tilde{a}^\dagger_k\ket{0^{(k)}_\beta},
\end{equation}
where $e^{-\beta\omega_k/2}$ is parametrized by the $\beta$-dependent thermal-mixing angle $\theta_k(\beta)$ via the relation\cite{khanna2009}
\begin{equation}
e^{-\beta\omega_k/2}
\equiv
\dfrac{\sinh\theta_k(\beta)}{\cosh\theta_k(\beta)}
=
\tanh\theta_k(\beta)
\end{equation}
leading to the definition of $\theta_k(\beta)$ in Eq.\eqref{eq.harmonic_tqp_mixing_angle} and the TBT given by Eqs.\eqref{eq.physical_tqp_operator} and \eqref{eq.auxiliary_tqp_operator}. The inverse TBT is given by the relations
\begin{align}
\hat{a}_k
&=
\cosh\theta_k(\beta)\,\hat{b}_k
+
\sinh\theta_k(\beta)\,\tilde{b}^\dagger_k,
\label{eq.inverse_tbt_phys}
\vspace{0.2em}
\\
\tilde{a}_k
&=
\cosh\theta_k(\beta)\,\tilde{b}_k
+
\sinh\theta_k(\beta)\,\hat{b}^\dagger_k.
\label{eq.inverse_tbt_aux}
\end{align}
Further, general TQP number states are obtained via $\hat{b}^\dagger_k$ and $\tilde{b}^\dagger_k$ acting on the multi-mode TQP thermal vacuum state, $\ket{\underline{0}_\beta}=\ket{0^{(1)}_\beta}\dots\ket{0^{(f)}_\beta}$, as
\begin{equation}
\ket{n^{(1)}_\beta,\dots,n^{(f)}_\beta,\tilde{m}^{(1)}_\beta,\dots,\tilde{m}^{(f)}_\beta}
=
\prod^f_{k,k^\prime=1}
\dfrac{\left(\hat{b}^\dagger_k\right)^{n_k}}{\sqrt{n_k!}}
\dfrac{\left(\tilde{b}^\dagger_{k^\prime}\right)^{m_{k^\prime}}}{\sqrt{m_{k^\prime}!}}
\underbrace{
\ket{0^{(1)}_\beta}
\dots
\ket{0^{(f)}_\beta}}_{=\ket{\underline{0}_\beta}}
\label{eq.thermal_tqp_numberstates}
\end{equation}
in analogy to the number state introduced in Eq.\eqref{eq.thermal_numberstates_definition}. The TQP number states together with $\ket{\underline{0}_\beta}$ also provide an orthonormal basis of the thermal Fock space $\mathcal{H}_\beta$. }

\section*{Appendix B: Numerical Parameters for Pyrazine Models}
The numerical parameters for the linear and bilinear 2-state-4-mode pyrazine models are reproduced from Ref.\cite{raabworth1999}. The linear non-adibatic coupling constant is given by $c_{10a}=0.208\,\text{eV}$ and the vertical energy gap by $\Delta=0.423\,\text{eV}$.
\begin{table}[hbt]
    \caption{Harmonic frequencies $\omega_k$, linear $a^{(1)}_k,a^{(2)}_k$ and quadratic $a^{(1)}_{kk},a^{(2)}_{kk}$ coupling constants for electronic states $\ket{S_1}$ and $\ket{S_2}$ as well as bilinear non-adiabatic coupling constants $c_{10a,k}$. All values are given in eV.
    }
    \vspace{0.25cm}
    \centering
    \begin{tabular}{c c c c c c c c c c}
       \hline
        mode $k$ \hspace{1.5em}          & $\omega_k$\hspace{1.5em}  && $a^{(1)}_k$\hspace{1.5em}  &&  $a^{(2)}_k$\hspace{1.5em}  &&  $a^{(1)}_{kk},\,a^{(2)}_{kk}$\hspace{1.5em} && $c_{10a,k}$ \\ 
       \hline\hline
       \vspace{0.2em}
       $v10a$\hspace{1.5em}    & 0.1139 &\hspace{1.5em}& 0.0 &\hspace{1em} & 0.0 &\hspace{1em} & -0.0116 &\hspace{1em}& 0.0\vspace{0.2em}\\ 
       $v6a$\hspace{1.5em}  & 0.0739 &\hspace{1.5em}& 0.0981 &\hspace{1em} & -0.1355 &\hspace{1em} & 0.0 &\hspace{1em}& 0.0055\vspace{0.2em}\\
       $v1$\hspace{1.5em}  & 0.1258  &\hspace{1.5em}& 0.503 &\hspace{1em}& 0.171 &\hspace{1em}& 0.0 &\hspace{1em}& 0.0100\vspace{0.2em}\\
       $v9a$\hspace{1.5em}  & 0.1525 &\hspace{1.5em}& 0.1452 &\hspace{1em}& 0.0375 &\hspace{1em}& 0.0 &\hspace{1em}& 0.0013\vspace{0.3em}\\
       \hline
    \end{tabular}
\label{tab.parameters_pyr_m4_lin}
\end{table}

\begin{table}[hbt]
    \caption{Bilinear intra-state coupling constants $a^{(1)}_{kk^\prime},a^{(2)}_{kk^\prime}$ for electronic states $\ket{S_1}$ and $\ket{S_2}$. All values are given in eV.
    }
    \vspace{0.5cm}
    \centering
    \begin{tabular}{c c c c c c }
       \hline
        $a^{(1)}_{kk^\prime}$\hspace{1.5em}   & $v6a$\hspace{1.5em}  && $v1$\hspace{1.5em}  &&  $v9a$\hspace{1.5em}  \vspace{0.15em}\\ 
       \hline\hline
       \vspace{0.2em}
       $v6a$\hspace{1.5em}  & 0.0 &\hspace{1.5em}& 0.0011 &\hspace{1em}& 0.002 \vspace{0.2em}\\
       $v1$\hspace{1.5em}  & 0.0011  &\hspace{1.5em}& 0.0 &\hspace{1em}& -0.0047 \vspace{0.2em}\\
       $v9a$\hspace{1.5em}  & 0.002 &\hspace{1.5em}& -0.0047 &\hspace{1em}& 0.0 \vspace{0.3em}\\
       \hline
    \end{tabular}
\hspace{1.5em}
    \begin{tabular}{c c c c c c }
       \hline
        $a^{(2)}_{kk}$\hspace{1.5em} & $v6a$ && $v1$ && $v9a$\vspace{0.3em}\\ 
       \hline\hline
       \vspace{0.2em}
       $v6a$\hspace{1.5em}  & 0.0 &\hspace{1.5em}& -0.0029 &\hspace{1em} & 0.0019 \vspace{0.2em}\\
       $v1$\hspace{1.5em}  & -0.0029  &\hspace{1.5em}& 0.0 &\hspace{1em}& -0.0016\vspace{0.2em}\\
       $v9a$\hspace{1.5em}  & 0.0019 &\hspace{1.5em}& -0.0016 &\hspace{1em}& 0.0\vspace{0.3em}\\
       \hline
    \end{tabular}
\label{tab.parameters_pyr_m4_bilin}
\end{table}

\section*{Appendix C: The Pyrazine Model Hamiltonian in TQP-TFD}
{We give explicit expressions for the vibronic coupling model Hamiltonian of pyrazine in thermal quasi-particle TFD. The diagonal contributions are given by
\begin{align}
\bar{H}^{(0)}_\beta
&=
\sum_{k=10a,6a,9a,1}
\hbar\omega_k
\left(
\hat{b}^\dagger_k
\hat{b}_k
-
\tilde{b}^\dagger_k
\tilde{b}_k
\right),
\vspace{0.2em}
\\
H^{(1)}_{\beta,i}
&=
\sum_{k=6a,9a,1}
\frac{a^{(i)}_k}{\sqrt{2}}
\biggl[
\cosh\theta_k
\left(
\hat{b}^\dagger_k
+
\hat{b}_k
\right)
+
\sinh\theta_k
\left(
\tilde{b}^\dagger_k
+
\tilde{b}_k
\right)
\biggr],
\vspace{0.2em}
\\
H^{(2)}_{\beta,i}
&=
\sum_{k,k^\prime=6a,9a,1}
\frac{a^{(i)}_{kk^\prime}}{2}
\left[
\cosh\theta_k\cosh\theta_{k^\prime}
\left(
\hat{b}^\dagger_k
+
\hat{b}_k
\right)
\left(
\hat{b}^\dagger_{k^\prime}
+
\hat{b}_{k^\prime}
\right)
\right.
\nonumber
\vspace{0.2em}
\\
&
\left.
+
\cosh\theta_k\sinh\theta_{k^\prime}
\left(
\hat{b}^\dagger_k
+
\hat{b}_k
\right)
\left(
\tilde{b}^\dagger_{k^\prime}
+
\tilde{b}_{k^\prime}
\right)
\right.
\nonumber
\vspace{0.4em}
\\
&
\left.
\hspace{2cm}
+
\sinh\theta_k\cosh\theta_{k^\prime}
\left(
\tilde{b}^\dagger_k
+
\tilde{b}_k
\right)
\left(
\hat{b}^\dagger_{k^\prime}
+
\hat{b}_{k^\prime}
\right)
\right.
\nonumber
\vspace{0.4em}
\\
&
\left.
\hspace{3.5cm}
+
\sinh\theta_k\sinh\theta_{k^\prime}
\left(
\tilde{b}^\dagger_k
+
\tilde{b}_k
\right)
\left(
\tilde{b}^\dagger_{k^\prime}
+
\tilde{b}_{k^\prime}
\right)
\right],
\end{align}
whereas the linear and quadratic vibronic coupling terms read
\begin{align}
\hat{V}^{(1)}_\beta
&=
\dfrac{c_{10a}}{\sqrt{2}}
\biggl[
\cosh\theta_{10a}
\left(
\hat{b}^\dagger_{10a}
+
\hat{b}_{10a}
\right)
+
\sinh\theta_{10a}
\left(
\tilde{b}^\dagger_{10a}
+
\tilde{b}_{10a}
\right)
\biggr],
\end{align}

\begin{align}
\hat{V}^{(2)}_\beta
&=
\sum_{k=6a,9a,1}
\frac{c_{10a,k}}{2}
\biggl[
\cosh\theta_{10a}\cosh\theta_k
\left(
\hat{b}^\dagger_{10a}
+
\hat{b}_{10a}
\right)
\left(
\hat{b}^\dagger_k
+
\hat{b}_k
\right)
\biggl.
\nonumber
\vspace{0.2cm}
\\
&
\biggl.
+
\cosh\theta_{10a}\sinh\theta_k
\left(
\hat{b}^\dagger_{10a}
+
\hat{b}_{10a}
\right)
\left(
\tilde{b}^\dagger_k
+
\tilde{b}_k
\right)
\biggr.
\nonumber
\vspace{0.2em}
\\
&
\biggl.
\hspace{2cm}
+
\sinh\theta_{10a}\cosh\theta_k
\left(
\tilde{b}^\dagger_{10a}
+
\tilde{b}_{10a}
\right)
\left(
\hat{b}^\dagger_k
+
\hat{b}_k
\right)
\biggr.
\nonumber
\vspace{0.2em}
\\
&
\biggl.
\hspace{3.5cm}
+
\sinh\theta_{10a}\sinh\theta_k
\left(
\tilde{b}^\dagger_{10a}
+
\tilde{b}_{10a}
\right)
\left(
\tilde{b}^\dagger_k
+
\tilde{b}_k
\right)
\biggr].
\end{align}}
For the bath, we additionally have
\begin{align}
\bar{H}^{(B)}_\beta
&=
\sum^{20}_{k=1}
\hbar\omega_{b,k}
\left(
\hat{b}^\dagger_{b,k}
\hat{b}_{b,k}
-
\tilde{b}^\dagger_{b,k}
\tilde{b}_{b,k}
\right),
\vspace{0.2cm}
\\
H^{(SB)}_{\beta,i}
&=
\sum_{k=1}^{20}
\dfrac{\kappa^{(i)}_k}{\sqrt{2}}
\biggl[
\cosh\theta_{b,k}
\left(
\hat{b}^\dagger_{b,k}
+
\hat{b}_{b,k}
\right)
+
\sinh\theta_{b,k}
\left(
\tilde{b}^\dagger_{b,k}
+
\tilde{b}_{b,k}
\right)
\biggr] \quad .
\end{align}

\section*{Appendix D: Thermal Ensemble Averages for Non-Adiabatic Dynamics in TFD} 
In general, we consider physical operators $\hat{O}=\hat{O}_v\otimes\hat{O}_e$ acting on the vibronic Hilbert space $\mathcal{H}_v\otimes\mathcal{H}_e$. The thermal average of a physical operator $\hat{O}$ is defined as
\begin{equation}
\braket{\hat{O}}_\beta(t)
=
\mathrm{tr}
\{
\hat{\rho}(t)\,
\hat{O}
\},
\label{eq.a_1}
\end{equation}
and, accordingly, the trace runs over both electronic and vibrational subspaces $\mathrm{tr}\{\dots\}=\mathrm{tr}_v\mathrm{tr}_e\{\dots\}$. Explicitly, we consider vibronic operators of the form
\begin{equation}
\hat{O}_1
=
\hat{1}_v\otimes\ket{S_i}\bra{S_i},
\hspace{0.5cm}
\hat{O}_2
=
\hat{n}_k
\otimes
\sum^2_{i=1}
\ket{S_i}\bra{S_i},
\end{equation}
with $\hat{n}_k=\hat{a}^\dagger_k\hat{a}_k$, to evaluate the diabatic populations via $\hat{O}_1$ and vibrational mean occupation number for the $k^{\text{th}}$ mode via $\hat{O}_2$. Starting with the diabatic populations, we have in TFD
\begin{align}
P^{(S_i)}_\beta(t)
&=
\mathrm{tr}
\{
\hat{\rho}(t)\,
\ket{S_i}\bra{S_i}
\}
\vspace{0.2cm}
\\
&=
\mathrm{tr}
\{
\mathrm{tr}_{\mathcal{\tilde{H}}}
\{
\ket{\Psi_\beta(t)}
\bra{\Psi_\beta(t)}
\}\,
\ket{S_i}\bra{S_i}
\}
\vspace{0.2cm}
\\
&=
\braket{
\Psi_\beta(t)
\vert
\biggl(
\ket{S_i}\bra{S_i}
\biggr)
\vert
\Psi_\beta(t)},
\end{align}
where we employed the cyclic invariance of the trace in the last line. For the vibrational mean occupation number, we have 
\begin{align}
\braket{\hat{n}_k}_\beta(t)
=
\mathrm{tr}
\{
\hat{\rho}(t)\,
\hat{a}^\dagger_k
\hat{a}_k
\}
=
\braket{
\Psi_\beta(t)
\vert
\hat{a}^\dagger_k
\hat{a}_k
\vert
\Psi_\beta(t)}
\end{align}
where we followed the same reasoning as for $P^{(S_i)}_\beta(t)$. In order to evaluate $\braket{\hat{n}_k}_\beta(t)$, the number operator  $\hat{n}_k=\hat{a}^\dagger_k\hat{a}_k$ is transformed to the TQP representation via the inverse TBT relations in Eqs.\eqref{eq.inverse_tbt_phys} and \eqref{eq.inverse_tbt_aux}, respectively. There, we find
\begin{align}
\hat{a}^\dagger_k\hat{a}_k
&=
\left(
\cosh\theta_k\,\hat{b}^\dagger_k
+
\sinh\theta_k\,\tilde{b}_k
\right)
\left(
\cosh\theta_k\,\hat{b}_k
+
\sinh\theta_k\,\tilde{b}^\dagger_k
\right),
\vspace{0.2cm}
\\
&=
\cosh^2\theta_k\,
\hat{b}^\dagger_k
\hat{b}_k
+
\sinh^2\theta_k\,
\tilde{b}_k
\tilde{b}^\dagger_k
+
\cosh\theta_k
\left(
\hat{b}^\dagger_k
\tilde{b}^\dagger_k
+
\hat{b}_k
\tilde{b}_k
\right),
\vspace{0.2cm}
\\
&=
\cosh^2\theta_k\,
\hat{b}^\dagger_k
\hat{b}_k
+
\sinh^2\theta_k\,
\tilde{b}^\dagger_k
\tilde{b}_k
+
\cosh\theta_k
\left(
\hat{b}^\dagger_k
\tilde{b}^\dagger_k
+
\hat{b}_k
\tilde{b}_k
\right)
+
\sinh^2\theta_k,
\end{align}
where we employed the commutation relation, $\tilde{b}_k\tilde{b}^\dagger_k=\tilde{b}^\dagger_k\tilde{b}_k+1$, in the third line. Employing the identity $\sinh^2\theta_k=\bar{n}_k(\beta)$, we finally obtain
\begin{multline}
\braket{\hat{n}_k}_\beta(t)
=
\cosh^2\theta_k
\braket{\hat{\mathfrak{n}}_k}_\beta(t)
+
\sinh^2\theta_k
\braket{\tilde{\mathfrak{n}}_k}_\beta(t)
\\
+
\cosh\theta_k\sinh\theta_k
\left(
\braket{\tilde{b}^\dagger_k\hat{b}^\dagger_k}_\beta(t)
+
\braket{\tilde{b}_k\hat{b}_k}_\beta(t)
\right)
+
\bar{n}_k(\beta),
\label{eq.tfd_vib_pop}
\end{multline}
with $\braket{\hat{\mathfrak{n}}_k}_\beta(t)=\braket{\hat{b}^\dagger_k\hat{b}_k}_\beta(t)$ and $\braket{\tilde{\mathfrak{n}}_k}_\beta(t)=\braket{\tilde{b}^\dagger_k\tilde{b}_k}_\beta(t)$, respectively. Here, $\bar{n}_k(\beta)=\left(e^{\beta\hbar\omega_k}-1\right)^{-1}$ is the Bose-Einstein distribution of the $k^{\text{th}}$ normal mode and $\braket{\hat{n}_k(t_0)}_\beta=\bar{n}_k(\beta)$, \textit{i.e.}, every vibrational mode is initially in a thermal equilibrium state.

\section*{Appendix E: Linear Absorption Spectra in TFD} 
In order derive the equivalence in Eq.\eqref{eq.tfd_acf}, we consider the full 3-state-4-mode Hamiltonian corresponding to Eq.\eqref{eq.pyrazine_system_4d_tqp} augmented by the diabatic electronic ground state 
\begin{equation}
\bar{H}_\beta
=
\left(
E_0
+
\bar{H}^{(0)}_\beta
\right)
\ket{S_0}
\bra{S_0}
+
\underbrace{\sum^2_{i=1}
\left(
E_i
+
\bar{H}^{(0)}_\beta
+
H^{(1)}_{\beta,i}
+
H^{(2)}_{\beta,i}
\right)
\ket{S_i}
\bra{S_i}
+
V_\beta
\biggl(
\ket{S_1}
\bra{S_2}
+
\ket{S_2}
\bra{S_1}
\biggr)}_{=\bar{H}^{(S)}_\beta}
\label{eq.tfd_acf_eq_a1}
\end{equation}
with $E_0=0$ and $E_2-E_1=2\Delta$ and excited state subspace Hamiltonian $\bar{H}^{(S)}_\beta$ identical to Eq.\eqref{eq.pyrazine_system_4d_tqp}. For the TFD autocorrelation function, one has 
\begin{align}
C_\beta(t)
&=
\braket{
\Psi_\beta
\vert
e^{-\text{i}\bar{H}_\beta\,t/\hbar}
\vert
\Psi_\beta},
\nonumber
\vspace{0.2cm}
\\
&=
\bra{S_2}
\braket{
\underline{0}_\beta
\vert 
e^{-\text{i}\bar{H}_\beta\,t/\hbar}
\vert
\underline{0}_\beta}
\ket{S_2},
\label{eq.tfd_acf_eq_a2}
\end{align}
with $\ket{\Psi_\beta}=\ket{S_2}\ket{\underline{0}_\beta}$. The initial vertical electronic excitation manifests via the action of the dipole operator in the Franck-Condon approximation as
\begin{equation}
\ket{S_2}
=
\biggl(
\underbrace{\mu_{20}
\left(
\ket{S_2}\bra{S_0}+\ket{S_0}\bra{S_2}
\right)
}_{=\hat{\mu}}
\biggr)\ket{S_0}
\label{eq.tfd_acf_eq_a3}
\end{equation}
with $\mu_{20}=1$ in the following. Inserting the latter into Eq.\eqref{eq.tfd_acf_eq_a2} leads to 
\begin{align}
C_\beta(t)
&=
\bra{S_0}
\bra{\underline{0}_\beta}
\biggl(
\hat{\mu}\,
e^{-\text{i}\bar{H}_\beta\,t/\hbar}
\hat{\mu}\,
\biggr)
\ket{\underline{0}_\beta}
\ket{S_0},
\nonumber
\vspace{0.2em}
\\
&=
\bra{S_0}
\bra{\underline{0}_\beta}
\biggl(
e^{\text{i}\bar{H}_\beta\,t/\hbar}
\hat{\mu}\,
e^{-\text{i}\bar{H}_\beta\,t/\hbar}
\hat{\mu}\,
\biggr)
\ket{\underline{0}_\beta}
\ket{S_0}.
\label{eq.tfd_acf_eq_a4}
\end{align}
In the second line, we used the relation $\bra{S_0}\bra{\underline{0}_\beta}e^{\text{i}\bar{H}_\beta\,t/\hbar}=\bra{S_0}\bra{\underline{0}_\beta}$ since with Eq.\eqref{eq.tfd_acf_eq_a1}, we have
\begin{equation}
\bra{S_0}\bra{\underline{0}_\beta}
\biggl(
\bar{H}^{(0)}_\beta
\,
\ket{S_0}\bra{S_0}
+
\bar{H}^{(S)}_\beta
\biggr)
=
0
\label{eq.tfd_acf_eq_a5}
\end{equation}
due to the orthogonal diabatic states and $\bra{\underline{0}_\beta}\bar{H}^{(0)}_\beta=0$. We now reverse the TBT, \textit{i.e.}, $\bar{H}_\beta=\bar{H}$ (\textit{cf.} Eqs. \eqref{eq.inverse_tbt_phys} and \eqref{eq.inverse_tbt_aux}), and find with $e^{-\text{i}\bar{H}\,t/\hbar}=e^{-\text{i}\hat{H}\,t/\hbar}e^{-\text{i}\tilde{H}\,t/\hbar}$ that
\begin{align}
C_\beta(t)
&=
\bra{S_0}
\bra{\underline{0}_\beta}
\biggl(
e^{\text{i}\bar{H}\,t/\hbar}
\hat{\mu}\,
e^{-\text{i}\bar{H}\,t/\hbar}
\hat{\mu}\,
\biggr)
\ket{\underline{0}_\beta}
\ket{S_0},
\nonumber
\vspace{0.2cm}
\\
&=
\bra{S_0}
\bra{\underline{0}_\beta}
\biggl(
e^{\text{i}\hat{H}\,t/\hbar}
\hat{\mu}\,
e^{-\text{i}\hat{H}\,t/\hbar}
\hat{\mu}\,
\biggr)
\ket{\underline{0}_\beta}
\ket{S_0},
\label{eq.tfd_acf_eq_a6}
\end{align}
as $\left[\hat{H},\tilde{H}\right]=0$ and $\left[\hat{\mu},\tilde{H}\right]=0$. We now perform traces with respect to the physical and auxiliary vibrational mode subspaces as well as the electronic subspace, which leads to
\begin{align}
C_\beta(t)
&=
\mathrm{tr}_e
\mathrm{tr}_v
\left\{
\mathrm{tr}_{\tilde{\mathcal{H}}}
\left\{
\bra{S_0}
\bra{\underline{0}_\beta}
\biggl(
e^{\text{i}\hat{H}\,t/\hbar}
\hat{\mu}\,
e^{-\text{i}\hat{H}\,t/\hbar}
\hat{\mu}\,
\biggr)
\ket{\underline{0}_\beta}
\ket{S_0}
\right\}
\right\},
\nonumber
\vspace{0.2cm}
\\
&=
\mathrm{tr}
\left\{
\mathrm{tr}_{\tilde{\mathcal{H}}}
\left\{
\biggl(
e^{\text{i}\hat{H}\,t/\hbar}
\hat{\mu}\,
e^{-\text{i}\hat{H}\,t/\hbar}
\hat{\mu}\,
\biggr)
\ket{\underline{0}_\beta}
\bra{\underline{0}_\beta}
\ket{S_0}
\bra{S_0}
\right\}
\right\},
\label{eq.tfd_acf_eq_a7}
\end{align}
where we wrote $\mathrm{tr}\{\dots\}=\mathrm{tr}_e\mathrm{tr}_v\{\dots\}$ and employed the cyclic invariance of the trace in the second line. We now note that only $\ket{\underline{0}_\beta}\bra{\underline{0}_\beta}$ involves contributions from auxiliary states in $\tilde{\mathcal{H}}_v$, such that the trace $\mathrm{tr}_{\tilde{\mathcal{H}}_v}\{\dots\}$ acts exclusively on this contribution. Therefore, it follows that
\begin{align}
C_\beta(t)
&=
\mathrm{tr}
\left\{
\biggl(
e^{\text{i}\hat{H}\,t/\hbar}
\hat{\mu}\,
e^{-\text{i}\hat{H}\,t/\hbar}
\hat{\mu}\,
\biggr)
\mathrm{tr}_{\tilde{\mathcal{H}}_v}
\{\ket{\underline{0}_\beta}\bra{\underline{0}_\beta}\}
\ket{S_0}\bra{S_0}
\right\},
\nonumber
\vspace{0.2em}
\\
&=
\mathrm{tr}
\left\{
\biggl(
e^{\text{i}\hat{H}\,t/\hbar}
\hat{\mu}\,
e^{-\text{i}\hat{H}\,t/\hbar}
\hat{\mu}\,
\biggr)
\hat{\rho}^0_\beta
\ket{S_0}\bra{S_0}
\right\},
\label{eq.tfd_acf_eq_a8}
\end{align}
where in the second line, we have $\mathrm{tr}_{\tilde{\mathcal{H}}_v}\left\{\ket{\underline{0}_\beta}\bra{\underline{0}_\beta}\}\right\}=\hat{\rho}^0_\beta$ by definition. Finally, we set $\hat{\rho}^0_\beta\ket{S_0}\bra{S_0}=\hat{\rho}(t_0)$ leading to
\begin{equation*}
C_\beta(t)
=
\mathrm{tr}
\left\{
\biggl(
e^{\text{i}\hat{H}\,t/\hbar}
\hat{\mu}\,
e^{-\text{i}\hat{H}\,t/\hbar}
\hat{\mu}\,
\biggr)
\hat{\rho}(t_0)
\right\},
\label{eq.tfd_acf_eq_a9}
\end{equation*}
which is the desired result.

\section*{Appendix F: MCTDH-TQP Expansion for (4+4)D Pyrazine Models}
We give numbers of tSPFs (\textit{cf.} Tabs.\ref{tab.parameters_mctdh_tqp_m4_lin} and \ref{tab.parameters_mctdh_tqp_m4_bilin}) and primitive TQP basis functions for MCTDH-TQP expansions, $T^{(2)}_0$, of thermofield states discussed in Tabs.\ref{tab.pyr_m4_linear_cpu} and \ref{tab.pyr_m4_bilinear_cpu} for the linear and bilinear (4+4)D pyrazine models. 

\begin{table}[h]
    \caption{Number of tSPFs for MCTDH-TQP runs of the linear 2-state-4-mode pyrazine model at different temperatures with combined modes in the format tSPF$(S_1)$/tSPF$(S_2)$.
    }
    \vspace{0.2cm}
    \centering
    \begin{tabular}{c c c c c c c c}
       \hline
        mode $k$\hspace{1.5em}   & $1\text{K}$\hspace{1.5em}  && $100\text{K}$\hspace{1.5em}  &&  $300\text{K}$\hspace{1.5em} &&  $500\text{K}$\hspace{1.5em}  \vspace{0.3em}\\ 
       \hline\hline
       \vspace{0.2em}
       $v10a,v6a$\hspace{1.5em}  & 19/11 &\hspace{1.5em}& 19/11 &\hspace{1em}& 21/13 &\hspace{1em}& 28/20\vspace{0.2em}\\
       $v9a,v1$\hspace{1.5em}  & 13/8 &\hspace{1.5em}& 13/8 &\hspace{1em}& 15/10 &\hspace{1em}& 20/15\vspace{0.2em}\\
       $tv10a,tv6a$\hspace{1.5em}  & 5/3  &\hspace{1.5em}& 8/5 &\hspace{1em}& 12/8 &\hspace{1em}& 23/15\vspace{0.2em}\\
       $tv9a,tv1$\hspace{1.5em}  & 3/2 &\hspace{1.5em}& 6/3 &\hspace{1em}& 8/5 &\hspace{1em}& 10/8\vspace{0.3em}\\
       \hline
    \end{tabular}
\label{tab.parameters_mctdh_tqp_m4_lin}
\end{table}

\begin{table}[h]
 \caption{Number of tSPFs for MCTDH-TQP runs of the bilinear 2-state-4-mode pyrazine model at different temperatures with combined modes in the format tSPF$(S_1)$/tSPF$(S_2)$.
    }
    \vspace{0.5cm}
    \centering
    \begin{tabular}{c c c c c c c c}
       \hline
        mode $k$\hspace{1.5em}   & $1\text{K}$\hspace{1.5em}  && $100\text{K}$\hspace{1.5em}  &&  $300\text{K}$\hspace{1.5em} &&  $500\text{K}$\hspace{1.5em}  \vspace{0.3em}\\ 
       \hline\hline
       \vspace{0.2em}
       $v10a,v6a$\hspace{1.5em}  & 19/15 &\hspace{1.5em}& 19/17 &\hspace{1em}& 25/23 &\hspace{1em}& 37/32\vspace{0.2em}\\
       $v9a,v1$\hspace{1.5em}  & 15/13 &\hspace{1.5em}& 15/13 &\hspace{1em}& 21/19 &\hspace{1em}& 25/23\vspace{0.2em}\\
       $tv10a,tv6a$\hspace{1.5em}  & 10/8 &\hspace{1.5em}& 10/8 &\hspace{1em}& 15/12 &\hspace{1em}& 25/22\vspace{0.2em}\\
       $tv9a,tv1$\hspace{1.5em}  & 10/8 &\hspace{1.5em}& 10/8 &\hspace{1em}& 19/15 &\hspace{1em}& 19/15\vspace{0.3em}\\
       \hline
    \end{tabular}
\label{tab.parameters_mctdh_tqp_m4_bilin}
\end{table}

In the MCTDH-SQR framework of the \textit{Heidelberg MCTDH package}\cite{heidelbergmctdh} a sine-DVR is employed for the primitive basis. In this work, we used $45\,(v10a)$, $35\,(v6a)$, $21\,(v1)$ and $12\,(v9a)$ primitive basis functions for the physical modes and $45\,(tv10a)$, $35\,(tv6a)$, $21\,(tv1)$ and $12\,(tv9a)$ for the auxiliary modes.

\end{document}